 \documentclass[aps,prd,twocolumn,superscriptaddress,showpacs]{revtex4}
\usepackage{graphicx}
\usepackage{amsmath}

\renewcommand{\bar}[1]{\overline{#1}}
\renewcommand{\bar}[1]{\overline{#1}}
\newcommand{\half}{{\frac{1}{2}}}

\newcommand{\ket}[1]{\,\left|\,{#1}\right\rangle}
\newcommand{\mbf}[1]{\mathbf{#1}}

\def\Dslash{\raise.15ex\hbox{/}\kern-.7em D}
\def\Pslash{\raise.15ex\hbox{/}\kern-.7em P}

\begin{document}

\preprint{SLAC-PUB-12544}

\title{Light-Front Dynamics and AdS/QCD Correspondence: 
The Pion Form Factor in the Space- and Time-Like Regions
}

\author{Stanley J. Brodsky}
\affiliation{Stanford Linear Accelerator Center, Stanford
University, Stanford, California 94309, USA}

\author{Guy F. de T\'eramond}
\affiliation{Universidad de Costa Rica, San Jos\'e, Costa Rica}
\affiliation{Stanford Linear Accelerator Center, Stanford
University, Stanford, California 94309, USA}

\date{\today}

\begin{abstract}

The AdS/CFT correspondence between string theory in AdS space and
conformal field theories in physical space-time leads to an
analytic, semi-classical model for strongly-coupled QCD which has
scale invariance and dimensional counting at short distances and
color confinement at large distances.  The
AdS/CFT correspondence also provides insights into the inherently
non-perturbative aspects of QCD such as the orbital and radial
spectra of hadrons and the form of hadronic wavefunctions.  In
particular, we show that there is an exact correspondence between
the fifth-dimensional coordinate of anti--de Sitter (AdS) space $z$ and a specific
light-front impact variable $\zeta$ which measures the separation of the quark
and gluonic constituents within the hadron in ordinary space-time.
This connection allows one to compute the analytic form of the
frame-independent light-front wavefunctions of mesons and baryons,
the fundamental entities which encode hadron properties and which
allow the computation of decay constants, form factors and other
exclusive scattering amplitudes.  
Relativistic light-front equations in ordinary space-time are found
which reproduce the results obtained using the fifth-dimensional
theory.   As specific examples we compute
the pion  coupling constant $f_\pi$, the pion charge radius 
$\left\langle r_\pi^2 \right\rangle$ and examine the propagation of the
electromagnetic current in AdS space, which determines the space and time-like behavior 
of the pion form factor and the pole of the $\rho$ meson.

\end{abstract}

\pacs{11.15.Tk, 11.25Tq, 12.38Aw, 12.40Yx}

\maketitle

\section{Introduction}

Quantum chromodynamics, the Yang-Mills local gauge field theory of
$SU(3)_C$ color symmetry provides a fundamental description of
hadron and nuclear physics in terms of quark and gluon degrees of
freedom. 
However, because of its strong coupling nature, it is
difficult to find analytic solutions to QCD or to make precise
predictions outside of its perturbative domain.  An important goal
is thus to find an initial approximation to QCD which is both
analytically tractable and which can be systematically improved. For
example, in quantum electrodynamics, the Coulombic Schr\"odinger and
Dirac equations provide quite accurate first approximations to
atomic bound state problems, which can then be systematically
improved using the Bethe-Salpeter formalism and correcting for
quantum fluctuations, such as the Lamb Shift and vacuum
polarization.

One of the most significant theoretical advances in recent years has
been the application of the AdS/CFT
correspondence~\cite{Maldacena:1997re} between string states defined
on the 5-dimensional anti--de Sitter (AdS) space-time and conformal
field theories in physical space-time. The essential principle
underlying the AdS/CFT approach to conformal gauge theories is the
isomorphism of the group of Poincare' and conformal transformations
$SO(4,2)$ to the group of isometries of anti-de Sitter space.  The
AdS metric is
\begin{equation} \label{eq:AdSz}
ds^2 = \frac{R^2}{z^2}(\eta_{\mu \nu} dx^\mu
 dx^\nu - dz^2),
 \end{equation}
which is invariant under scale changes of the
coordinate in the fifth dimension $z \to \lambda z$ and $ x_\mu \to
\lambda x_\mu$.  Thus one can match scale transformations of the
theory in $3+1$ physical space-time to scale transformations in the
fifth dimension $z$.

QCD is not itself a conformal theory; however as we shall discuss, 
there are indications, both from
theory and phenomenology, that the QCD coupling is slowly varying at small momentum transfer.
Thus in the domain where
the QCD coupling is approximately constant and quark masses can be
neglected, QCD resembles a conformal theory. As shown by Polchinski
and Strassler~\cite{Polchinski:2001tt}, one can simulate confinement
by imposing boundary conditions in the holographic variable at
$z = z_0 ={1/\Lambda_{\rm QCD}}$. Confinement can also be introduced
by modifying the AdS metric to mimic a confining potential.
The resulting models, although {\it ad hoc}, provide a
simple semi-classical approximation to QCD which has both constituent counting
rule behavior at short distances and confinement at large distances.
This simple approach, which has been described as a ``bottom-up''
approach, has been successful in obtaining
general properties of scattering amplitudes of hadronic bound
states at strong coupling~\cite{Janik:1999zk, Polchinski:2001tt, Brodsky:2003px, Brower:2007qh}, the low-lying hadron
spectra~\cite{Boschi-Filho:2002vd, deTeramond:2004qd, deTeramond:2005su,
Erlich:2005qh, Karch:2006pv, deTeramond:2006xb, Evans:2006ea, Hong:2006ta, Colangelo:2007pt, Forkel:2007cm}, 
hadron couplings and chiral symmetry breaking~\cite{Erlich:2005qh, 
DaRold:2005zs, Hirn:2005nr, Ghoroku:2005vt},
quark potentials in confining
backgrounds~\cite{Boschi-Filho:2005mw, Andreev:2006ct} and
a description of weak hadron decays ~\cite{Hambye:2005up}. 
Geometry back-reaction in AdS may also be relevant to the 
infrared physics~\cite{Csaki:2006ji}.
The gauge theory/gravity duality also provides a convenient framework
for the description of deep inelastic scattering structure functions at small $x$~\cite{Polchinski:2002jw},
a unified description of hard and soft pomeron physics~\cite{Brower:2006ea} and
gluon scattering amplitudes at strong coupling~\cite{Alday:2007hr}.
Recent applications to describe chiral symmetry breaking~\cite{Babington:2003vm} and other meson and baryon properties, have also been carried out within the framework of a top-bottom approach to AdS/CFT using higher dimensional branes~\cite{Kirsch:2006he, Apreda:2006bu,
Nawa:2006gv, Hong:2007kx, Hata:2007mb}. 

The AdS/QCD correspondence is particularly relevant for the description
of hadronic form factors,  since it incorporates the connection between the twist of the hadron to the fall-off of its current matrix elements, as well as  essential aspects of vector meson dominance.
It also provides a convenient framework for analytically continuing the space-like results to the time-like region. Recent applications to the form factors of
mesons and nucleons~\cite{deTeramond:2006xb, Radyushkin:2006iz, Brodsky:2006ys, Grigoryan:2007vg, Grigoryan:2007my, Brodsky:2007pt, Choi:2006ha}
have followed from the original work described
in~\cite{Polchinski:2002jw, Hong:2004sa}

A physical hadron in four-dimensional Minkowski space has
four-momentum $P_\mu$ and 
invariant mass  given by $P_\mu P^\mu = \mathcal{M}^2$. 
The physical states in  AdS$_5$ space are represented by normalizable ``string" modes
\begin{equation} \label{eq:hadronicmode}
\Phi_P(x,z) \sim  e^{- i P\cdot x}~ \Phi(z),
\end{equation}
with plane waves along the Poincar\'e coordinates and a profile
function $\Phi(z)$ along the holographic coordinate $z$. 
For small-$z$, $\Phi$ scales as $\Phi \sim z^\Delta$, 
where $\Delta$ is the conformal dimension of the string state, the
same dimension of the interpolating operator $\mathcal{O}$ which creates a hadron
out of the vacuum~\cite{Polchinski:2001tt}, 
$\langle P \vert \mathcal{O} \vert 0 \rangle \ne 0$.
The scale dependence of each 
string mode $\Phi(x,z)$ is thus determined by matching its behavior at $z \to 0$
with the dimension of 
the corresponding hadronic state at short distances $x^2 \to 0$.
Changes in length scale are mapped to evolution in the holographic
variable $z$. The string mode  
in $z$ thus represents the extension
of the hadron wave function into the fifth dimension.
There are also non-normalizable modes which are related to 
external currents: they propagate into the interior of AdS space. 

As we shall discuss, there is a remarkable mapping between the AdS description of hadrons and the Hamiltonian formulation of QCD in physical space-time quantized on the light front. 
The light-front wavefunctions of bound states in QCD are relativistic and frame-independent
generalizations of the familiar Schr\"odinger wavefunctions of
atomic physics, but they are determined at fixed light-cone time
$\tau  = t +z/c$---the ``front form" advocated by Dirac---rather than
at fixed ordinary time $t$.  An important feature of light-front
quantization is the fact that it provides exact formulas to write
down  matrix elements as a sum of bilinear forms, which can be mapped
into their AdS/CFT counterparts in the semi-classical approximation.
One can thus obtain not only an
accurate description of the hadron spectrum for light quarks, but also
a remarkably simple but realistic model of the valence wavefunctions
of mesons, baryons, and glueballs.  In fact, we find an exact correspondence between the
fifth-dimensional coordinate of anti-de Sitter space $z$ and a
specific impact variable $\zeta$ in the light-front formalism which
measures the separation of the constituents within the hadron in
ordinary space-time.  Since $\tau = 0$ in the front  form, it corresponds
to the transverse separation $x^\mu x_\mu = - \mbf{x}_\perp^2$.
The amplitude $\Phi(z)$ describing  the
hadronic state in $\rm{AdS}_5$ can then be precisely mapped to the
light-front wavefunctions $\psi_{n/h}$ of hadrons in physical
space-time~\cite{Brodsky:2006uq}, thus providing a relativistic
description of hadrons in QCD at the amplitude level. 

This paper is organized as follows.  We first review in section II recent evidence which indicates that QCD has  near-conformal  behavior in the infrared region.   The QCD light-front Fock representation is then briefly reviewed in section III. In section IV, we derive the exact form of current matrix elements in the light-front formalism which sets the framework for the identification with the corresponding amplitudes in AdS/CFT, and constitutes the basis of the partonic interpretation of the AdS/CFT correspondence. 
The actual mapping to AdS matrix elements is carried out in section V for the hard wall model and in section VI for the soft wall model, where the truncated space boundary conditions are replaced by a smooth cutoff.   
Results for the pion form factor in the space and time-like regions and the pion charge radius
$\langle r_\pi^2 \rangle$ are presented in section VII for both
models. In section VIII we compute the pion decay constant in light-front QCD and find its precise expression in AdS, following the mapping discussed in sections V and VI. Some final remarks are given in the conclusions in section IX. Other technical aspects useful for the discussion of the article are given in the appendices. In particular we
relate our results for the soft wall model to the analytical results found recently by Grigoryan and 
Radyushkin~\cite{Grigoryan:2007my}.

\section{Conformal QCD Window}

It was originally believed that the AdS/CFT mathematical correspondence could
only be applied to strictly conformal theories, such as
$\mathcal{N}=4$ supersymmetric Yang-Mills gauge theory.   
In our approach, we will apply the gauge theory/gravity duality to the strong coupling 
regime of QCD where the coupling appears to be approximately constant.  
Theoretical studies based on Dyson-Schwinger equation~\cite{vonSmekal:1997is,  Zwanziger:2003cf,  Alkofer:2004it, Epple:2006hv}   and
phenomenological~\cite{Deur:2005cf, Brodsky:2002nb} evidence is in fact
accumulating that the QCD couplings defined from physical observables such
as $\tau$ decay~\cite{Brodsky:1998ua} become constant at small
virtuality; {\em i.e.}, effective charges develop an infrared fixed
point in contradiction to the usual assumption of singular growth in
the infrared.  Recent lattice gauge theory
simulations~\cite{Furui:2006py}  also indicate an
infrared fixed point for QCD.     The near-constant behavior of effective couplings thus suggests that
QCD can be approximated as a conformal theory at relatively
small momentum transfer.

It is clear from a physical perspective that 
in a confining theory where gluons and quarks have an effective mass or maximal wavelength, all
vacuum polarization corrections to the gluon self-energy must decouple at
long wavelength; thus an infrared fixed point appears to be a natural consequence of confinement. 
Furthermore, if one considers a
semi-classical approximation to QCD with massless quarks and without
particle creation or absorption, then the resulting $\beta$ function
is zero, the coupling is constant, and the approximate theory is
scale and conformal invariant.

In the case of hard exclusive reactions~\cite{Lepage:1980fj}, the virtuality of the gluons exchanged in the underlying QCD process is typically much less than the momentum transfer scale $Q$, since typically several gluons share the total momentum transfer.  Since the coupling is probed in the conformal window, this kinematic feature can explain why the measured  proton Dirac form factor scales as $Q^4 F_1(Q^2) \simeq {\rm const}$ up to $Q^2 < 35$ GeV$^2$~\cite{Diehl:2004cx} with little sign of the logarithmic running of the QCD coupling.

Thus conformal symmetry can be a useful
first approximant even for physical QCD.
Conformal symmetry is broken in physical QCD by quantum effects and quark masses;
nevertheless, one can use conformal symmetry as a {\it template},
systematically correcting for its nonzero $\beta$ function as well
as higher-twist effects. For example, ``commensurate scale
relations"~\cite{Brodsky:1994eh} which relate QCD observables to each
other, such as the generalized Crewther
relation~\cite{Brodsky:1995tb}, have no renormalization scale or
scheme ambiguity and retain a convergent perturbative structure
which reflects the underlying conformal symmetry of the classical
theory.  In general, the scale is set such that one has the correct
analytic behavior at the heavy particle
thresholds~\cite{Brodsky:1982gc}.  Analytic
effective charges~\cite{Brodsky:1998mf} such as the pinch
scheme~\cite{Binger:2006sj,Cornwall:1989gv}  also provide an important perspective for unifying the
electroweak and strong couplings~\cite{Binger:2003by}.

\section{The Light-Front Fock Representation}

The light-front expansion of any hadronic system
is constructed by quantizing quantum chromodynamics
at fixed light-front time \cite{Dirac:1949cp} $\tau = t + z/c$.
In terms of the hadron  four-momentum $P = (P^+, P^-, \mbf{P}_{\!\perp})$,
$P^\pm = P^0 \pm P^3$,
the light-front Lorentz invariant Hamiltonian for the composite system, 
$H_{LF}^{QCD} = P^-P^+ - \mbf{P}^2_\perp$,  has
eigenvalues given in terms of the eigenmass ${\cal M}$ squared  corresponding 
to the mass spectrum of the color-singlet states in QCD~\cite{Brodsky:1997de}.
The momentum generators $P^+$ and
$\mbf{P}_{\! \perp}$ are kinematical; {\em i.e.}, they are independent of
the interactions. The LF time evolution operator  
$P^- = i \frac{d}{d\tau}$ 
can be derived directly from the QCD Lagrangian in the light-front
gauge $A^+=0$.
In principle, the complete set of bound state and scattering
eigensolutions of $H_{LF}$ can be obtained by solving the light-front
Heisenberg equation
\begin{equation} \label{eq:HLC}
H_{LF} \ket{\psi_h} = {\cal M}^2_h \ket{\psi_h},
\end{equation}
where $\ket{\psi_h}$ is an expansion in multi-particle Fock eigenstates
$\{\ket{n} \}$ of the free light-front
Hamiltonian: 
$\vert \psi_h \rangle = \sum_n \psi_{n/h} \vert \psi_h \rangle $.
The LF Heisenberg
equation has in fact been solved for QCD$(1\! + \!1)$ and a number of
other theories using the discretized light-cone quantization method~\cite{Pauli:1985ps}.
The light-front gauge
has the advantage that all gluon degrees of freedom have physical
polarization and positive metric.  In addition,  orbital angular
momentum has a simple physical interpretation in this
representation. The light-front wavefunctions (LFWFs) $\psi_{n/h}$ provide a
frame independent representation of a hadron, and relate the quark
and gluon degrees of freedom with their asymptotic hadronic state.

Given the light-front wavefunctions $\psi_{n/h}$ one can 
compute a large range of hadron
observables. For example, the valence and sea quark and gluon
distributions which are measured in deep inelastic lepton scattering
are defined from the squares of the LFWFs summed over all Fock
states $n$. Form factors, exclusive weak transition
amplitudes~\cite{Brodsky:1998hn} such as $B\to \ell \nu \pi$,  and
the generalized parton distributions~\cite{Brodsky:2000xy} measured
in deeply virtual Compton scattering are (assuming the ``handbag"
approximation) overlaps of the initial and final LFWFs with $n
=n^\prime$ and $n =n^\prime+2$. The gauge-invariant distribution
amplitude $\phi_H(x_i,Q)$ defined from the integral over the
transverse momenta $\mbf{k}^2_{\perp i} \le Q^2$ of the valence
(smallest $n$) Fock state provides a fundamental measure of the
hadron at the amplitude level~\cite{Lepage:1979zb, Efremov:1979qk};
they  are the nonperturbative input to the factorized form of hard
exclusive amplitudes and exclusive heavy hadron decays in
perturbative QCD. The resulting distributions obey the 
Dokshitzer-Gribov-Altarelli-Parisi (DGLAP) and
Efremov-Radyushkin-Brodsky-Lepage
(ERBL) evolution equations as a function of the maximal invariant
mass, thus providing a physical factorization
scheme~\cite{Lepage:1980fj}. In each case, the derived quantities
satisfy the appropriate operator product expansions, sum rules, and
evolution equations. At large $x$ where the struck quark is
far-off shell, DGLAP evolution is quenched~\cite{Brodsky:1979qm}, so
that the fall-off of the deep inelastic scattering (DIS) cross sections in $Q^2$ satisfies
inclusive-exclusive duality at fixed $W^2.$

The hadron wavefunction is an eigenstate of the total momentum $P^+$
and $\mbf{P}_{\! \perp}$ and the longitudinal spin projection $S_z$,
and is normalized according to
\begin{multline}
\bigl\langle \psi_h(P^+,\mbf{P}_{\! \perp}, S_z) \big\vert 
\psi_h(P'^+,\mbf{P}'_\perp,S_z') \bigr\rangle \\ 
= 2 P^+ (2 \pi)^3 \,\delta_{S_z,S'_z} \,\delta \bigl(P^+ - P'^+ \bigr)
\,\delta^{(2)} \negthinspace \bigl(\mbf{P}_{\! \perp} - \mbf{P}'_\perp\bigr) . 
\label{eq:Pnorm}
\end{multline}

Each hadronic eigenstate $\vert \psi_h \rangle$  is expanded in
a Fock-state complete basis of non-interacting $n$-particle states
$\vert n \rangle$ with an infinite number of components
\begin{multline} 
\left\vert \psi_h(P^+,\mbf{P}_{\! \perp}, S_z) \right\rangle = \\
\sum_{n,\lambda_i} 
\prod_{i=1}^n \int \! \frac{dx_i d^2 \mbf{k}_{\perp i}}{2 \sqrt{x_i} (2\pi)^3} \, (16 \pi^3) \,
\delta \Bigl(1 - \sum_{j=1}^n x_j\Bigr) 
\delta^{(2)} \negthinspace\Bigl(\sum_{j=1}^n\mbf{k}_{\perp j}\Bigr) \\ \times
\psi_{n/h}(x_i,\mbf{k}_{\perp i},\lambda_i) 
\bigl\vert n: x_i P^+\!, x_i \mbf{P}_{\! \perp} \! + \! \mbf{k}_{\perp i},\lambda_i \bigr\rangle,
\label{eq:LFWFexp}
\end{multline}
where the sum begins with the valence state; e.g., $n \ge 2$ for mesons. The
coefficients of the  Fock expansion
\begin{equation} \label{eq:LFWF}
\psi_{n/h}(x_i, \mbf{k}_{\perp i},\lambda_i) 
= \bigl\langle n:x_i,\mbf{k}_{\perp i},\lambda_i \big\vert \psi_h\bigr\rangle ,
\end{equation}
are independent of the total momentum $P^+$ and $\mbf{P}_{\! \perp}$ of
the hadron and depend only on the relative partonic coordinates,
the longitudinal momentum fraction $x_i = k_i^+/P^+$,
the relative transverse momentum $\mbf{k}_{\perp i}$,
and $\lambda_i$, the
projection of the constituent's spin along the $z$ direction. 
Thus, given the Fock-projection (\ref{eq:LFWF}), the wavefunction
of a hadron is determined in any frame. The
amplitudes $\psi_{n/h}$ represent the probability amplitudes to find
on-mass-shell constituents $i$ with longitudinal momentum $ x_i P^+$, 
transverse momentum $x_i \mbf{P}_{\! \perp} + \mbf{k}_{\perp i}$,
helicity $\lambda_i$ and invariant mass
\begin{equation}
\mathcal{M}_n^2 = \sum_{i=1}^n k_i^\mu k_{i \mu} = \sum_{i=1}^n
\frac{\mbf{k}^2_{\perp i} + m_i^2}{x_i},
\end{equation}
in the hadron $h$. Momentum conservation requires
$\sum_{i=1}^n x_i =1$ and $\sum_{i=1}^n \mbf{k}_{\perp i}=0$.
In addition, each light front wavefunction
$\psi_{n/h}(x_i,\mbf{k}_{\perp i},\lambda_i)$ obeys the angular momentum sum 
rule~\cite{Brodsky:2000ii}
\begin{equation}
J^z  = \sum_{i=1}^n  S^z_i + \sum_{i=1}^{n-1} L^z_i ,
 \end{equation}
where $S^z_i = \lambda_i $ and the $n-1$ orbital angular momenta
have the operator form 
\begin{equation}
L^z_i =-i \left(\frac{\partial}{\partial k^x_i}k^y_i -
\frac{\partial}{\partial k^y_i}k^x_i \right). 
\end{equation}
The LFWFs are normalized according to
\begin{equation}
\sum_n  \int \big[d x_i\big] \left[d^2 \mbf{k}_{\perp i}\right]
\,\left\vert \psi_{n/h}(x_i, \mbf{k}_{\perp i}) \right\vert^2 = 1,
\label{eq:LFWFnorm}
\end{equation}
where the measure of the constituents phase-space momentum
integration  is
\begin{equation}
\int \big[d x_i\big] \equiv
\prod_{i=1}^n \int dx_i \,\delta \Bigl(1 - \sum_{j=1}^n x_j\Bigr) ,
\end{equation}
\begin{equation}
\int \left[d^2 \mbf{k}_{\perp i}\right] \equiv \prod_{i=1}^n \int
\frac{d^2 \mbf{k}_{\perp i}}{2 (2\pi)^3} \, (16 \pi^3) \,
\delta^{(2)} \negthinspace\Bigl(\sum_{j=1}^n\mbf{k}_{\perp j}\Bigr),
\end{equation}
for the normalization given by (\ref{eq:Pnorm}). 
The spin indices have been suppressed.

\subsection{Construction of Partonic States}

The complete basis of Fock-states $\ket{n}$ is constructed by applying 
free-field creation operators to
the vacuum state $\vert 0 \rangle$ which has no particle content,
$P^+ \vert 0 \rangle =0$, $\mbf{P}_{\! \perp} \vert 0 \rangle = 0$. The
basic constituents appear from the light-front quantization of the
excitations of the dynamical fields, the Dirac field $\psi_+$,
$\psi_\pm = \Lambda_\pm \psi$, $\Lambda_\pm = \gamma^0 \gamma^\pm$,
and the transverse field $\mbf{A}_\perp$ in the $A^+ = 0$ gauge,
expanded in terms of creation and annihilation operators on the
transverse plane with coordinates $x^- = x^0 - x^3$ and $\mbf{x}_\perp$ 
at fixed light-front time $x^+ = x^0 +
x^3$~\cite{Brodsky:1997de};
\begin{multline} \label{eq:psiop}
\psi_+(x)_\alpha = \sum_\lambda \int_{q^+ > 0} \frac{d q^+}{\sqrt{ 2
 q^+}}
\frac{d^2 \mbf{q}_\perp}{ (2 \pi)^3} \\ \times
\left[b_\lambda (q)
u_\alpha(q,\lambda) e^{-i q \cdot x} + d_\lambda (q)^\dagger
v_\alpha(q,\lambda) e^{i q \cdot x}\right],
\end{multline}
\begin{multline}
\mbf{A}_\perp(x) = \sum_\lambda \int_{q^+ > 0} 
\frac{d q^+}{\sqrt{2
 q^+}} \frac{d^2 \mbf{q}_\perp}{(2 \pi)^3} \\  \times
\left[a(q)
\vec\epsilon_\perp(q,\lambda) e^{-i q \cdot x} + a(q)^\dagger
\vec\epsilon_\perp^{\,*}(q,\lambda) e^{i q \cdot x } \right],
\end{multline}
with commutation relations
\begin{multline} \label{eq:cr}
\left[a(q), a^\dagger(q')\right] = \left\{b(q),
b^\dagger(q')\right\} = \left\{d(q), d^\dagger(q')\right\} = \\
(2
 \pi)^3 \,\delta (q^+ - {q'}^+) ~
\delta^{(2)}\negthinspace\left(\mbf{q}_\perp - \mbf{q}'_\perp\right).
\end{multline}
We shall use the conventions of Appendix A of Ref. \cite{Lepage:1980fj} for the
properties of the light-cone spinors.
A one-particle state is defined by
\begin{equation} \label{eq:cop}
\vert q \rangle = \sqrt{2 q^+} \,b^\dagger(q) \vert 0 \rangle,
\end{equation}
so that its normalization has the Lorentz invariant form
\begin{equation} \label{eq:normq}
\langle q | q' \rangle =  2 {q}^+  (2 \pi)^3  \delta ({q}^+ -
{q'}^+)\, \delta^{(2)} (\mbf{q}_{\perp} - \mbf{q}_{\perp}'),
\end{equation}
and this fixes our normalization. Each n-particle Fock state
$|p_i^+, \mbf{p}_{\perp i} \rangle$ is an eigenstate of  $P^+$ and
$\mbf{P}_{\! \perp}$ and is normalized according to
\begin{multline}
\bigl\langle  p_i^+, \mbf{p}_{\perp i},\lambda
 \big|{p'}_i^+, \mbf{p'}_{\negthinspace\perp i},\lambda' \bigr\rangle
= \\ 2 p_i^+ (2 \pi)^3 \, \delta \bigl(p_i^+ - {p'}_i^+\bigr) \, 
\delta^{(2)} \negthinspace \bigl(\mbf{p}_{\perp i} - \mbf{p'}_{\negthinspace\perp
    i}\bigr) \, \delta_{\lambda,\lambda'}.
\label{eq:normFC}
\end{multline}

\subsection{Impact Space Representation}

The holographic mapping of hadronic LFWFs to AdS string modes
is most transparent when one uses the impact parameter space representation.
The total position coordinate of a hadron or its transverse center
of momentum $\mbf{R}_\perp$, is defined in terms of the energy
momentum tensor $T^{\mu \nu}$
\begin{equation}
\mbf{R}_\perp = \frac{1}{P^+} \int dx^- 
\negthinspace \int d^2 \mbf{x}_\perp \,T^{++} \,
\mbf{x}_\perp.
\end{equation}
In terms of partonic transverse coordinates
\begin{equation}
x_i \mbf{r}_{\perp i} = x_i \mbf{R}_\perp + \mbf{b}_{\perp i},
\end{equation}
where  the $\mbf{r}_{\perp i}$ are the physical transverse position
coordinates and  $\mbf{b}_{\perp i}$ frame independent  internal
coordinates, conjugate to the relative coordinates $\mbf{k}_{\perp i}$. 
Thus, $\sum_{i=1}^n \mbf{b}_{\perp i} = 0$ and  
$\mbf{R}_\perp = \sum_{i=1}^n x_i \mbf{r}_{\perp i}$.
The LFWFs $\psi_n(x_j, \mbf{k}_{\perp j})$ can be expanded in terms of the $n-1$ independent
transverse coordinates $\mbf{b}_{\perp j}$,  $j = 1,2,\dots,n-1$
\begin{multline} \label{eq:LFWFb}
\psi_n(x_j, \mbf{k}_{\perp j}) =  (4 \pi)^{(n-1)/2} 
\prod_{j=1}^{n-1}\int d^2 \mbf{b}_{\perp j} \\
\exp{\Big(i \sum_{j=1}^{n-1} \mbf{b}_{\perp j} \cdot \mbf{k}_{\perp j}\Big)} \,
\tilde{\psi}_n(x_j, \mathbf{b}_{\perp j}).
\end{multline}
The normalization is defined by
\begin{equation}  
\sum_n  \prod_{j=1}^{n-1} \int d x_j d^2 \mbf{b}_{\perp j} 
\left\vert\tilde \psi_n(x_j, \mbf{b}_{\perp j})\right\vert^2 = 1.
\end{equation}

\section{The Form Factor in QCD}

The hadron form factors can be predicted from the overlap integral
of string modes propagating in AdS space with the boundary electromagnetic
source which probes the AdS interior, or equivalently by using the 
Drell-Yan-West formula in physical space-time. If both quantities
represent the same physical observable for any value of the
transfer momentum $q^2$, an exact correspondence can be
established between the string modes $\Phi$ in fifth-dimensional
AdS space and the light-front wavefunctions of hadrons $\psi_{n/h}$
in 3+1 space-time~\cite{Brodsky:2006uq}.

\subsection{The Drell-Yan-West Form Factor}

The light-front formalism
provides an exact Lorentz-invariant representation
of current  matrix elements in terms of the 
overlap of light-front wave functions. 
The electromagnetic current has elementary couplings to the
charged constituents
since the full Heisenberg current can
be replaced by the free quark current $J^\mu(0)$, evaluated at fixed 
light-cone time $x^+=0$ in the $q^+=0$ frame~\cite{Drell:1969km}. 
In contrast to the covariant Bethe-Salpeter equation,
in the light front Fock expansion 
one does not need to include the contributions to the current from an 
infinite number of irreducible kernels, or the interactions
of the electromagnetic current with vacuum 
fluctuations~\cite{Brodsky:1997de}.

We choose the light-front frame coordinates
\begin{eqnarray} \label{eq:qframe}
P  &=& (P^+, P^-, \mbf{P}_{\! \perp}) = \Bigl( P^+,\frac{ M^2}{ P^+},
\vec 0_{\perp} \Bigr),\\q  &=& (q^+, q^-, \mbf{q}_{\perp}) = \Bigl(
0,  \frac{2 \,q \! \cdot \! P}{P^+}, \mbf{q}_{\perp} \Bigr), \nonumber
\end{eqnarray}
where  $q^2 = - Q^2 = -2 \,q \! \cdot \! P
 = - \mbf{q}^2_\perp$ is the space-like
four momentum squared transferred to the composite system. The
electromagnetic form factor of a meson is defined in terms of the
hadronic  amplitude of the electromagnetic current evaluated at
light-cone time $x^+ = 0$
\begin{equation}
\left\langle P' \left\vert J^+(0) \right\vert P\right\rangle = 2
\left( P + P'\right)^+ F(Q^2),
\end{equation}
where $P' = P + q$ and $F(0) = 1$.

The expression for the current operator $J^+(x) = \sum_q e_q
\bar\psi(x) \gamma^+ \psi(x)$ in the particle number representation
follows from the momentum expansion of $\psi(x)$ in terms of creation
and annihilation operators given by (\ref{eq:psiop}). Since $\gamma^+$ 
conserves the spin component of the struck quark (Appendix A),
we obtain for $J^+(0)$
\begin{multline} \label{eq:Jem}
J^+(0) =  \sum_q e_q
\int \frac{d q^+ d^2\mbf{q}_\perp}{(2 \pi)^3} 
\int \frac{d q'^+ d^2\mbf{q}'_\perp}{(2 \pi)^3} \\ \times
\bigl\{ b_\uparrow^\dagger(q) b_\uparrow(q')
     + b_\downarrow^\dagger(q)  b_\downarrow(q') 
   -  d_\uparrow^\dagger(q)  d_\uparrow(q')
     - d_\downarrow^\dagger(q)  d_\downarrow(q') \bigr\}.
\end{multline}
The operator $J$ annihilates a quark (antiquark) with with charge
$e_q$ ($-e_q$), momentum $q'$ and spin-up (spin-down) along the $z$
direction and creates a quark (antiquark) with the same spin and
momentum $q$.

The matrix element 
$\left\langle\psi_{P'}  \left \vert J^+(0) \right\vert \psi_P \right\rangle$ can be computed
by expanding the initial and final hadronic states
in terms of its Fock components using (\ref{eq:LFWFexp}). The transition
amplitude can then be expressed as a sum of overlap integrals with diagonal $J^+$-matrix
elements in the $n$-particle Fock-state basis.
For each Fock-state, we label with $q$ the struck
constituent quark with charge $e_q$ and $j = 1, 2 \dots, n-1$ each
spectator. Using the normalization condition
(\ref{eq:normFC}) for each individual constituent and after
integration over the intermediate variables in the $q^+ = 0$ frame,
we find the Drell-Yan-West (DYW) expression for the meson
form factor~\cite{Drell:1969km, West:1970av}
\begin{multline} \label{eq:DYW}
F(q^2) = \sum_n  \int \big[d x_i\big] \left[d^2 \mbf{k}_{\perp i}\right] \\ \times
\sum_j e_j \psi^*_{n/P'} (x_i, \mbf{k}'_{\perp i},\lambda_i)
\psi_{n/P} (x_i, \mbf{k}_{\perp i},\lambda_i),
\end{multline}
where the variables of the light-cone Fock components in the
final-state are given by $\mbf{k}'_{\perp i} = \mbf{k}_{\perp i} 
+ (1 - x_i)\, \mbf{q}_\perp $ for a struck  constituent quark and 
$\mbf{k}'_{\perp i} = \mbf{k}_{\perp i} - x_i \, \mbf{q}_\perp$ for each
spectator. The formula is exact if the sum is over all Fock states $n$. 
Notice that there is a factor of $N_C$ from a closed quark loop where the photon
is attached and a normalization factor of $1/\sqrt{N_C}$ for each
meson wave function; thus color factors cancel out from the
expression of the form factor.

\subsection{The Form Factor in Impact Space}

One of the important advantages of the light-front formalism is that current
matrix elements can be represented without approximation as overlaps
of light-front wavefunctions. In the case of the elastic space-like
form factors, the matrix element of the $J^+$ current only couples
Fock states with the same number of constituents, as expressed by
the Drell-Yan-West formula (\ref{eq:DYW}). 
Suppose that the charged parton $n$ is the active constituent struck by the
current, and the quarks $i = 1,2, \cdots ,n-1$ are spectators.
We substitute (\ref{eq:LFWFb}) in the DYW formula (\ref{eq:DYW}). 
Integration over $k_\perp$ phase space gives us $n - 1$ delta
functions to integrate over the $n - 1$ intermediate transverse variables
with the result
\begin{multline} \label{eq:FFb} 
F(q^2) =  \sum_n  \prod_{j=1}^{n-1}\int d x_j d^2 \mbf{b}_{\perp j} 
\exp \! {\Bigl(i \mbf{q}_\perp \cdot \sum_{j=1}^{n-1} x_j \mbf{b}_{\perp j}\Bigr)} \\ \times
\left\vert \tilde \psi_n(x_j, \mbf{b}_{\perp j})\right\vert^2,
\end{multline}
corresponding to a change of transverse momentum $x_j \mbf{q}_\perp$ for each
of the $n-1$ spectators. This is a convenient form for
comparison with  AdS results, since the form factor
may be expressed in terms of the product of light-front wave
functions  with identical variables.

\subsection{Effective  Single-Particle Distribution}

The form factor in the light-front formulation has an exact representation in terms of an effective single particle  
density~\cite{Soper:1976jc}
\begin{equation}
F(q^2) = \int_0^1 dx ~\rho(x, \mbf{q}_\perp),
\end{equation}
where $\rho(x, \mbf{q}_\perp)$ is given by
\begin{multline} \label{eq:rhoeta}
\rho(x, \mbf{q}_\perp) = \sum_n \prod_{j=1}^{n-1}
\int dx_j \, d^2 \mbf{b}_{\perp j} \, \delta \Bigl(1-x - \sum_{j=1}^{n-1} x_j\Bigr) \\ \times
\exp{\Bigl(i \mbf{q}_\perp
\cdot \sum_{j=1}^{n-1} x_j \mbf{b}_{\perp j}\Bigr)}
\Bigl| \tilde \psi_n(x_j, \mbf{b}_{\perp j})\Bigr|^2.
\end{multline}
The integration is over the coordinates of the $n-1$ spectator
partons, and  $x = x_n$ is the coordinate of the active charged
quark.  
We can also write the form factor in terms of an effective  single
particle transverse distribution $\tilde\rho(x,\vec \eta_\perp)$~\cite{Soper:1976jc}
\begin{equation} \label{eq:FFeta}
F(q^2) =\int^1_0 dx  \int d^2 \vec \eta_\perp e^{i \vec \eta_\perp \cdot \mbf{q}_\perp}
\tilde\rho(x,\vec \eta_\perp),
\end{equation}
where
$\vec \eta_\perp = \sum^{n-1}_{j=1} x_j \mbf{b}_{\perp j}$
is the $x$-weighted transverse position coordinate of the $n-1$
spectators~\cite{Burkardt:2003je}. The corresponding transverse density is
\begin{multline} \label{eq:rhoeta}
\tilde\rho(x,\vec \eta_\perp) = 
\int \frac{d^2 \mbf{q}_\perp}{(2 \pi)^2}  
e^{-i \vec \eta_\perp \cdot \mbf{q}_\perp} \rho(x, \mbf{q}_\perp) \\ 
= \sum_n \prod_{j=1}^{n-1} \int dx_j \, d^2\mbf{b}_{\perp j} 
\,\delta \Bigl(1-x-\sum_{j=1}^{n-1} x_j\Bigr) \\ \times
\delta^{(2)}\Bigl(\sum_{j=1}^{n-1} x_j \mbf{b}_{\perp j} - \vec \eta_\perp\Bigr)
\left\vert \tilde\psi_n(x_j, \mbf{b}_{\perp j}) \right\vert^2.
\end{multline}
The procedure is valid for any Fock state $n,$ and thus the results can be summed over 
$n$ to obtain an exact representation. 


\section{The Form Factor in AdS/CFT and Light-Front Mapping of
String Amplitudes}

We now derive the corresponding expression for the electromagnetic  form factor of a pion in
AdS. The derivation can
be extended to vector mesons and baryons, although the actual analysis becomes more complex,
since the matrix elements of space-like local operators include the dependence for the
different spin transitions. For example, in the case of the proton, the Dirac and Pauli form
factors correspond to the light-cone spin-conserving and spin-flip matrix elements of the
$J^+$ current~\cite{Brodsky:1980zm}, and a mapping to the corresponding AdS matrix elements has to be carried out for each case. The electric vector-meson form factor has been discussed recently
in the context of the hard wall model in~\cite{Hong:2004sa, Grigoryan:2007vg} and in the
soft-wall model in~\cite{Grigoryan:2007my}. 
A detailed discussion of the form factor of nucleons in AdS/QCD  will be given elsewhere.

In AdS/CFT, the hadronic matrix element for the electromagnetic current 
has the form of a convolution of the string modes for the initial and final
hadrons with the external electromagnetic source which propagates inside AdS.
We discuss first the truncated space or hard wall~\cite{Polchinski:2001tt} 
holographic model, where  quark and gluons as well as the external electromagnetic 
current propagate freely into the AdS interior according to
the AdS metric. We discuss later the case where the sharp boundary of the hard wall in the infrared
region is replaced by the introduction of a soft cutoff.

AdS coordinates are the Minkowski coordinates
$x^\mu$ and $z$, the holographic coordinate, which we label collectively 
$x^\ell = (x^\mu, z)$. The metric of the full 
space-time is
\begin{equation}
ds^2 = g_{\ell m} dx^\ell dx^m 
\label{eq:metric}
\end{equation}
where $g_{\ell m} \to \left(R^2/z^2\right) \eta_{\ell m}$
in the conformal  $z \to 0$ limit, and $\eta_{\ell m}$ has diagonal components
$(1, -1, \cdots, -1)$.
Unless stated otherwise, 
5-dimensional fields are represented by capital letters such as
$\Phi$. Holographic fields in 4-dimensional Minkowski space are represented by lower case
letters such as $\phi$. The metric (\ref{eq:metric}) is AdS
in the finite interval $0 \leq z \leq z_0 = 1/\Lambda_{QCD}$, and the 
dual conformal field theory is strongly coupled at all scales
up to the confining scale. 

Assuming minimal coupling the form factor
has the form~\cite{Polchinski:2002jw, Hong:2004sa}
\begin{equation}
i g_5 \int d^4x \, dz \,\sqrt{g}\, A^{\ell}(x,z)
\Phi_{P'}^*(x,z) \overleftrightarrow\partial_\ell \Phi_P(x,z),
\label{eq:M}
\end{equation} 
where $g_5$ is a five-dimensional effective coupling constant and
$\Phi_P(x,z)$ is a normalizable mode representing a hadronic state,
$\Phi_P(x,z) \sim e^{-iP \cdot x} \Phi(z)$, with hadronic invariant mass
given by $P_\mu P^\mu = \mathcal{M}^2$.

We consider the propagation inside AdS space of an electromagnetic
probe polarized along Minkowski coordinates  ($Q^2 = - q^2 > 0$)
\begin{equation}
A(x,z)_\mu = \epsilon_\mu e^{-i Q \cdot x} J(Q^2,z),~~~ A_z = 0,
\end{equation}
where $J(Q^2, z)$  has the value 1 at
zero momentum transfer, since we are normalizing the bulk solutions to the total charge operator,
and as boundary limit the external current 
$A_\mu(x, z \to 0) = \epsilon_\mu e^{-i Q \cdot x}$.
Thus 
\begin{equation} \label{eq:Jbc}
J(Q^2 = 0, z) = J(Q^2 ,z = 0) = 1.
\end{equation}

The propagation of the external current inside AdS space is described by the AdS
wave equation 
\begin{equation} \label{eq:AdSJ}
\left[ z^2 \partial_z^2 -  z \, \partial_z - z^2 Q^2 \right]   J(Q^2, z)  = 0,
 \end{equation}
and its solution subject to the boundary conditions (\ref{eq:Jbc}) is
\begin{equation} \label{eq:J}
J(Q^2, z) = z Q K_1(z Q).
\end{equation}
Notice that $J(Q^2, z)$  can also be obtained
from the Green's function of (\ref{eq:AdSJ}),
since it is the bulk-to-boundary propagator~\cite{Hong:2004sa, Erlich:2005qh}.

Substituting the normalizable mode $\Phi(x,z)$ in (\ref{eq:M}) and extracting an overall delta
function from momentum conservation at the vertex, we find for the elastic form factor of the pion
\begin{equation}
\left\langle P' \left\vert J^\mu(0) \right\vert P \right\rangle = 2
(P + P')^\mu F(Q^2),
\end{equation}
with
\begin{equation} 
F(Q^2)  =  R^3  \! \int \frac{dz}{z^3} \, \Phi(z) J(Q^2, z) \Phi(z).
\label{eq:FFAdS}
\end{equation}
The form factor in AdS is thus 
represented as the overlap in the fifth dimension coordinate $z$
of the normalizable modes dual to the incoming
and outgoing hadrons, $\Phi_P$ and $\Phi_{P'}$, with the
non-normalizable mode, $J(Q^2, z)$, dual to the external
source~\cite{Polchinski:2002jw}.
Since $K_n(x) \sim \sqrt{\pi/2 x} \, e^{-x}$ for large $x$, it follows
that the external electromagnetic field is suppressed inside the AdS cavity
for large $Q$. At small $z$ the string modes
scale as $\Phi \sim z^\Delta$.
At large enough $Q$, the important contribution to (\ref{eq:FFAdS})
is from the region near $z \sim 1/Q$ \begin{equation}
F(Q^2) \to \left[\frac{1}{Q^2}\right]^{\Delta - 1},
\end{equation}
and the ultraviolet point-like behavior~\cite{Polchinski:2001ju} responsible for the power law scaling~\cite{Brodsky:1973kr, Matveev:ra} is recovered. This is a remarkable consequence of truncating AdS 
space~\cite{Polchinski:2001tt}, since we are indeed describing the coupling of an electromagnetic
current to an extended mode, and we should expect soft collision amplitudes characteristic of strings, instead of hard point-like ultraviolet behavior~\cite{Polchinski:2001tt}.

\subsection{Light-Front Mapping in the Hard-Wall Model}

We can now establish a connection of the AdS/CFT and the LF formulae.
It is useful to integrate (\ref{eq:FFeta}) over angles to obtain
\begin{equation} \label{eq:FFzeta} 
F(q^2) = 2 \pi \int_0^1 dx \frac{(1-x)}{x}  \int \zeta d \zeta\,
J_0 \negthinspace \left(\negthinspace\zeta q \sqrt{\frac{1-x}{x}}\right) \tilde \rho(x,\zeta),
\end{equation}
where we have introduced the variable
\begin{equation}
\zeta = \sqrt{\frac{x}{1-x}} ~\Big\vert \sum_{j=1}^{n-1} x_j \mathbf{b}_{\perp j}\Big\vert,
\end{equation}
representing the $x$-weighted transverse impact coordinate of the
spectator system. We substitute the integral representation of the bulk-to
boundary propagator $J(Q^2, z)$ (Appendix B)
\begin{equation} \label{eq:intJ}
J(Q^2, z) = \int_0^1 \! dx \, J_0\negthinspace \left(\negthinspace\zeta Q
\sqrt{\frac{1-x}{x}}\right) = \zeta Q K_1(\zeta Q),
\end{equation}
 in the expression for the form factor in
AdS space (\ref{eq:FFAdS}) for arbitrary values of $Q$.
Comparing with the light-front expression (\ref{eq:FFzeta}),
 we can identify the spectator density
function appearing in the light-front 
formalism with the corresponding AdS density
\begin{equation} \label{eq:PhirhoHW}
\tilde \rho(x,\zeta)
=   \frac{R^3}{2 \pi} \frac{x}{1-x}
\frac{\left\vert \Phi(\zeta)\right\vert^2}{\zeta^4} .
\end{equation}
Equation (\ref{eq:PhirhoHW}) gives a precise relation between  string modes $\Phi(\zeta)$ 
in AdS$_5$ and the QCD transverse charge density $\tilde\rho(x,\zeta)$. 
The variable $\zeta$, $0 \leq \zeta \leq \Lambda_{\rm QCD}^{-1}$, represents a measure of the transverse
separation between point-like constituents, and it is also the
holographic variable $z$~\cite{Brodsky:2006uq}.

In the case of a two-parton constituent bound state the correspondence between the string 
amplitude $\Phi(z)$ in AdS space and the QCD light-front wave wavefunction 
 $\widetilde \psi\left(x, \mbf{b}_\perp\negthinspace\right)$
follows immediately from (\ref{eq:PhirhoHW}). For two partons the transverse density
(\ref{eq:rhoeta}) has the simple form
\begin{equation}
\tilde\rho_{n=2}(x, \zeta) = 
\frac{\vert \tilde\psi(x,\zeta)\vert^2}{(1-x)^2},
\end{equation}
and a closed form solution for the two-constituent bound state
light-front wave function is obtained
\begin{equation} \label{eq:Phipsi}
\vert \tilde\psi(x,\zeta)\vert^2 = 
\frac{R^3}{2 \pi} \, x(1-x)
\frac{\vert \Phi(\zeta)\vert^2}{\zeta^4},
\end{equation}
where $\zeta^2 =  x(1-x) \mathbf{b}_\perp^2$.
Here $\mbf{b}_\perp$ is the impact separation and Fourier conjugate to $\mbf{k}_\perp$.

\subsection{Holographic Light-Front Representation}

The mapping of the holographic variable $z$ from AdS space to the impact variable $\zeta$ in the LF frame allows
the equations of motion in AdS space to be recast in the form of  a
light-front Hamiltonian equation~\cite{Brodsky:1997de}
\begin{equation}
H_{LF} \! \ket{\phi} = \mathcal{M}^2 \! \ket{\phi},
\end{equation} 
a remarkable result which  allows the discussion of the AdS/CFT
solutions in terms of light-front equations in physical 3+1 space-time.  
To make more transparent the holographic connection
between AdS$_5$ and the conformal quantum field theory defined at
its asymptotic $z\to 0$ boundary, it is convenient to write  the AdS
metric (\ref{eq:AdSz}) in terms of light front coordinates $x^\pm =
x^0 \pm x^3$ and $\mbf{x}_\perp$
\begin{equation} \label{eq:AdSLF}
ds^2 = \frac{R^2}{z^2} \left( dx^+ dx^- - d \mbf{x}_\perp^2 - dz^2
\right).
\end{equation}

To simplify  the discussion we consider  the propagation of massive scalar modes in AdS space
described by the normalized solutions to the wave equation ($d=4$)
\begin{equation} 
\label{eq:AdSPhiM}
\left[z^2 \partial_z^2 - (d-1) z\,\partial_z + z^2 \mathcal{M}^2 
- (\mu R)^2 \right] \Phi(z) = 0,
\end{equation} 
where $\mu$ is the five-dimensional mass. Since the field $\Phi$ couples to a local boundary interpolating operator of dimension $\Delta$, $\mathcal{O}_\Delta$, $\mu$ is determined by the
relation $(\mu R)^2 = \Delta(\Delta-4)$ between the fifth-dimensional mass of $\Phi$ and its scaling dimension. For spin-carrying constituents, the dimension of the operator is replaced by its twist $\tau$,
dimension minus spin, $\tau = \Delta - \sigma$, $\sigma = \sum_{i =1}^n \sigma_i$. For example $\tau = 2$ for a meson.

By substituting
\begin{equation} \label{eq:scalarshift}
\phi(\zeta) = \left(\frac{\zeta}{R}\right)^{-3/2} \Phi(\zeta),
\end{equation}
in (\ref{eq:AdSPhiM})
we find an effective Schr\"odinger equation as a function of the
weighted impact variable $\zeta$
\begin{equation} \label{eq:Scheqphi} 
\left[-\frac{d^2}{d \zeta^2} + V(\zeta) \right] \phi(\zeta) = \mathcal{M}^2 \phi(\zeta),
\end{equation}
with $- \frac{d^2}{d\zeta^2}$ the light-front kinetic energy operator and
\begin{equation}
V(\zeta) = - \frac{1-4 L^2}{4\zeta^2},
\end{equation}
the effective potential in the conformal limit, where we identify $\zeta$ with the fifth
dimension $z$ of AdS space: $\zeta = z$. We have written above $(\mu R)^2 = - 4 + L^2$,
for the fifth dimensional mass $\mu$ appearing in the AdS wave equation.
The effective wave equation
(\ref{eq:Scheqphi}) is a relativistic light-front equation defined at
$x^+ = 0$. The AdS metric $ds^2$ (\ref{eq:AdSLF}) is invariant if
$\mbf{x}_\perp^2 \to \lambda^2 \mbf{x}_\perp^2$ and $z \to \lambda
z$ at equal light-front time. The Casimir operator for the rotation
group $SO(2)$ in the transverse light-front plane is $L^2$. This
shows the natural holographic connection to the light front.

The  Breitenlohner-Freedman stability bound~\cite{Breitenlohner:1982jf}
\begin{equation}
(\mu R)^2 \ge - \frac{d^2}{4},
\end{equation}
requires  $L^2 \ge 0$, thus the lowest state corresponds to $L = 0$.
Higher excitations are matched to the small $z$ asymptotic behavior of each string mode
to the corresponding conformal dimension of the boundary operators
of each hadronic state. In the semiclassical approximation the solution for mesons with 
relative orbital angular momentum $L$ corresponds to an effective five dimensional mass on the
bulk side. The allowed values of $\mu R$ are determined by
requiring that asymptotically the dimensions become spaced by
integers. The solution to (\ref{eq:Scheqphi}) is
\begin{equation}
\phi(\zeta) = \zeta^{-3/2} \Phi(\zeta) = C \zeta^{1/2} J_L\left(\zeta \mathcal{M}\right).
\end{equation}
The holographic light-front wavefunction
$\phi(\zeta) = \langle \zeta \vert \phi \rangle$ is normalized 
according to
\begin{equation}
\langle \phi \vert \phi \rangle =
\int d\zeta \, \vert \langle  \zeta \vert \phi \rangle \vert^2 = 1,
\end{equation}
and represent the probability
amplitude to find $n$-partons at transverse impact separation $\zeta =
z$. Furthermore, its eigenmodes determine the hadronic mass spectrum~\cite{Brodsky:2006uq}.
For the truncated space model the
eigenvalues are determined by the boundary conditions at 
$\phi(z = 1/\Lambda_{\rm QCD}) = 0$ and are 
given in terms of the roots of  Bessel functions: 
$\mathcal{M}_{L,k} = \beta_{L,k} \Lambda_{\rm QCD}$.~\cite{ft:NeuBC}

\begin{figure}[h]
\centering
\includegraphics[angle=0,width=8.65cm]{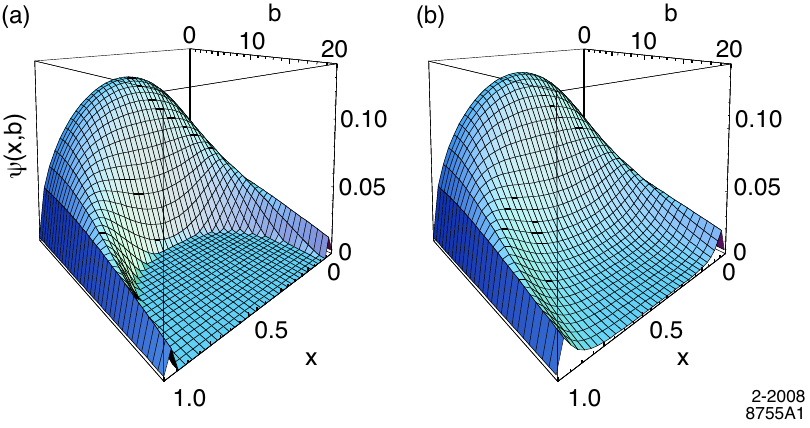}
\caption{AdS/QCD Predictions for the ground state light-front wavefunction of a pion in impact space. 
(a): truncated-space model for $\Lambda_{\rm QCD} = 0.22$ GeV. (b):  soft-wall model for $\kappa = 0.375$ GeV.}
\label{fig:MesonLFWF}
\end{figure}

A closed form of the light-front wavefunctions $\tilde\psi(x, \mbf{b}_\perp)$ in the hard
wall model follow from
(\ref{eq:Phipsi}) 
\begin{multline} 
\tilde \psi_{L,k}(x, \mbf{b}_\perp) 
=  \frac {\Lambda_{\rm QCD}}{\sqrt{ \pi} J_{1+L}(\beta_{L,k})} \sqrt{x(1-x)} \\ \times
J_L \! \left(\sqrt{x(1-x)} \, \vert\mbf{b}_\perp\vert \beta_{L,k} \Lambda_{\rm QCD}\right) 
\theta \! \left(\mbf{b}_\perp^2 \le \frac{\Lambda^{-2}_{\rm QCD}}{x(1-x)}\right).
\end{multline}
The resulting wavefunction depicted in Fig. \ref{fig:MesonLFWF} 
displays confinement at large interquark
separation and conformal symmetry at short distances, reproducing dimensional counting rules for hard exclusive amplitudes and the conformal properties of the LFWFs at high relative
momenta in agreement  with perturbative QCD.

\section{Soft-Wall Holographic Model}

The predicted mass spectrum in the truncated space hard-wall (HW) model is 
linear $M \propto L$ at high orbital
angular momentum $L$, in contrast to the quadratic dependence $M^2 \propto L$ in the usual
Regge parameterization. 
It has been shown recently that by  choosing a specific profile for a non-constant dilaton, the usual Regge  dependence  can be obtained~\cite{Karch:2006pv}. The procedure allows one
to retain conformal AdS metrics (\ref{eq:AdSz}) and to introduce a smooth cutoff  which depends on the profile of a dilaton background field $\varphi$ 
\begin{equation}
S = \int \! d^4x \, dz  \,\sqrt{g} \,e^{- \varphi(z) } \mathcal{L},
\end{equation}
where $\varphi$ is a function of the holographic coordinate $z$ which vanishes in the ultraviolet 
limit $z \to 0$. The IR hard-wall or truncated-space holographic model, discussed in the previous section, 
correspond to a constant dilaton field $\varphi(z) \! = \! \varphi_0$ in the confining region,  
$z \leq 1/\Lambda_{\rm QCD}$, and to very large values elsewhere: $\varphi(z) \to \infty$ for 
$z >  1/\Lambda_{\rm QCD}$. The introduction of a soft cutoff avoids the
ambiguities in the choice of boundary conditions at the infrared wall.  A convenient choice~\cite{Karch:2006pv} for the background field with
usual Regge behavior is $\varphi(z) = \kappa^2 z^2$.
The resulting wave equations are equivalent to the radial equation of a two-dimensional
oscillator, previously found in the context  of mode propagation on 
$AdS_5 \times S^5$, in the light-cone formulation of Type II supergravity~\cite{Metsaev:1999kb}.
Also,  equivalent results follow from
the introduction of a gaussian warp factor in the AdS metric
for the particular case of  massless vector modes propagating in the distorted 
metric~\cite{Andreev:2006vy}.

\subsection{Light-Front Mapping in the Soft-Wall Model}

We consider the propagation in AdS space of an electromagnetic probe polarized along Minkowski coordinates. Since the non-normalizable modes also couples to the dilaton field we must study the solutions of the modified wave equation
\begin{equation} \label{eq:AdSJSW}
\left[z^2 \partial_z^2   - z \left( 1 + 2 \kappa ^2 z^2   \right) \partial_z - Q^2 z^2 \right] 
J_\kappa(Q^2, z) = 0,
\end{equation}
subject to the boundary conditions (\ref{eq:Jbc}). The solution is
\begin{equation} \label{eq:Jkappa}
J_\kappa(Q^2, z) = \Gamma\!\left(1+ \frac{Q^2}{4 \kappa^2}\right) 
U\!\left(\frac{Q^2}{4 \kappa^2}, 0 , \kappa^2 z^2\right),
\end{equation}
where $U(a,b,c)$ is the confluent hypergeometric function.

The form factor in
AdS space in presence of the dilaton background $\varphi = \kappa^2 z^2$
has the additional term $e^{- \kappa^2 z^2}$  in the metric 
\begin{equation} 
F(Q^2) = R^3 \int \frac{dz}{z^3} \, e^{- \kappa^2 z^2} \Phi(z) J_\kappa(Q^2, z) \Phi(z),
\label{eq:FFAdSSW}
\end{equation}
to be  normalized to the charge operator at $Q = 0$.
In the large $Q^2$ limit, $Q^2 \gg 4 \kappa^2$, we find (Appendix C)
\begin{equation} \label{eq:Jasymp}
J_\kappa(Q^2, z) \to z Q K_1(z Q) = J(Q^2, z).
\end{equation}
Thus, for large transverse momentum the current decouples from the dilaton field, 
and we recover our previous scaling results for the ultraviolet behavior of matrix elements.

To obtain the corresponding basis set of LFWFs we compare the DYW expression of the form
factor  (\ref{eq:FFzeta}) with the AdS form factor (\ref{eq:FFAdSSW})
for large values  of $Q$, where (\ref{eq:Jasymp}) is valid.
Thus, in the large $Q$ limit we can identify the light-front spectator density
with the corresponding AdS density
\begin{equation} \label{eq:PhirhoSW}
\tilde \rho(x,\zeta)
=   \frac{R^3}{2 \pi} \frac{x}{1-x} \, e^{- \kappa^2 \zeta^2} \,
\frac{ \left\vert \Phi(\zeta)\right\vert^2}{\zeta^4} .
\end{equation}

When summed over all Fock states,  the Drell-Yan-West  formula gives an exact result. The
formula describes the coupling of the free electromagnetic current to the elementary constituents
in the interaction representation. In the presence of a dilaton field in AdS space, or in the case
where the
electromagnetic probe propagates in modified confining AdS metrics, the electromagnetic
AdS mode is no longer dual to a the free quark current, but  dual 
to a dressed current, {\it i.e.}, a hadronic electromagnetic current including virtual $\bar q q$
pairs and thus confined. Consequently, at finite values of the momentum transfer $Q^2$ our
simple identification discussed above has to be reinterpreted since we are comparing states in
different representations:  the interaction representation in light-cone QCD versus the Heisenberg representation in AdS. However both quantities should represent the same observables.
We thus expect that the modified mapping corresponds to the presence of higher Fock states in the
hadron.

\subsection{Holographic Light-Front Representation}

The equations of motion for scalar modes with smooth boundary conditions can also be
recast in the form of a light-cone Hamiltonian form $H_{LC} \! \ket{\phi} = \mathcal{M}^2 \! \ket{\phi}$,
with  effective potential
\begin{equation}  \label{eq:VSW}
V(\zeta) = - \frac{1-4 L^2}{4\zeta^2} + \kappa^4 \zeta^2 + 2 \kappa^2(L-1),
\end{equation}
in the covariant light-front effective Schr\"odinger equation (\ref{eq:Scheqphi}). 
The effective potential describes transverse oscillations in the light-front plane with SO(2)  rotation
subgroup and Casimir invariant $L^2$, representing LF rotations for the transverse coordinates.
The solution to equation (\ref{eq:Scheqphi}) for the potential (\ref{eq:VSW}) is
\begin{equation} \label{eq:phiLSW} 
\phi(\zeta) = \kappa^{1+L} \, \sqrt{\frac{2 n!}{(n\!+\!L\!)!}} \, \zeta^{1/2+L}
e^{- \kappa^2 \zeta^2/2} L^L_n(\kappa^2 \zeta^2),
\end{equation}
with eigenvalues
\begin{equation}
\mathcal{M}^2 = 4 \kappa^2 (n + L).
\end{equation}
The lowest stable solution corresponds to $n = L = 0$. 
With this choice for the potential, the lowest possible state is a zero mode in the spectrum, which can be identified with the pion. 

Upon the substitution 
\begin{equation}
\Phi(\zeta) = e^{ \kappa^2 \zeta^2/2} \left(\frac{\zeta}{R}\right)^{3/2} \phi(\zeta), ~~~ \zeta \to z,
\end{equation}
we express the solution (\ref{eq:phiLSW}) as the scalar mode
\begin{equation}
\Phi(z) = \frac{\kappa^{1+L}}{R^{(3/2)}} \, \sqrt{\frac{2 n!}{(n\!+\!L\!)!}} \, z^{2+L}
 L^L_n(\kappa^2 z^2),
\end{equation}
of scaling dimension $\Delta = 2 + L$ propagating in AdS space.
A closed form of the light-front wavefunctions $\widetilde\psi(x, \mbf{b}_\perp)$ for a two-parton state
in the soft wall model then follows  from (\ref{eq:PhirhoSW}) 
\begin{multline} 
 \tilde \psi(x,\mbf{b}_\perp)  = \frac{\kappa^{L+1}}{\sqrt{\pi}} 
 \sqrt{\frac{ n!}{(n+L)!}} \, [x(1-x)]^{\half+L} \vert \mbf{b}_\perp \vert^L  \\ \times
 e^{-\half \kappa^2 x(1-x) \mbf{b}_\perp^2} 
 L^L_n\big(\kappa^2 x(1-x) \mbf{b}_\perp^2\big).
 \end{multline}
 The resulting wavefunction for the pion is compared in Fig. \ref{fig:MesonLFWF} with the hard-wall 
result.

\section{Pion Form Factor}

We compute the pion form factor from the AdS expressions  (\ref{eq:FFAdS}) and (\ref{eq:FFAdSSW})
for the hadronic string modes $\Phi_\pi$ in the HW
\begin{equation} \label{eq:PhipiHW}
\Phi_\pi^{HW}(z) = \frac{\sqrt{2} \Lambda_{QCD}}{R^{3/2} J_1(\beta_{0,1}) } 
z^2 J_0\left(z \beta_{0,1} \Lambda_{QCD} \right) ,
\end{equation}
and soft-wall (SW) model
\begin{equation} \label{eq:PhipiSW}
\Phi_\pi^{SW}(z) = \frac{\sqrt{2} \kappa}{R^{3/2}}\, z^2 ,
\end{equation}
respectively. We have obtained numerical results for both models. 
It was found recently that in the case of the SW model 
the results for form factors can be expressed analytically~\cite{Grigoryan:2007my}.
Thus, we shall extend the analytical results given in~\cite{Grigoryan:2007my} to 
string modes with arbitrary scaling dimension $\tau$,
integer or not, allowing us to study the effect of 
anomalous dimensions in the expression for the form factors. This is described in Appendix D.

\begin{figure}[h]
\centering
\includegraphics[angle=0,width=8.0cm]{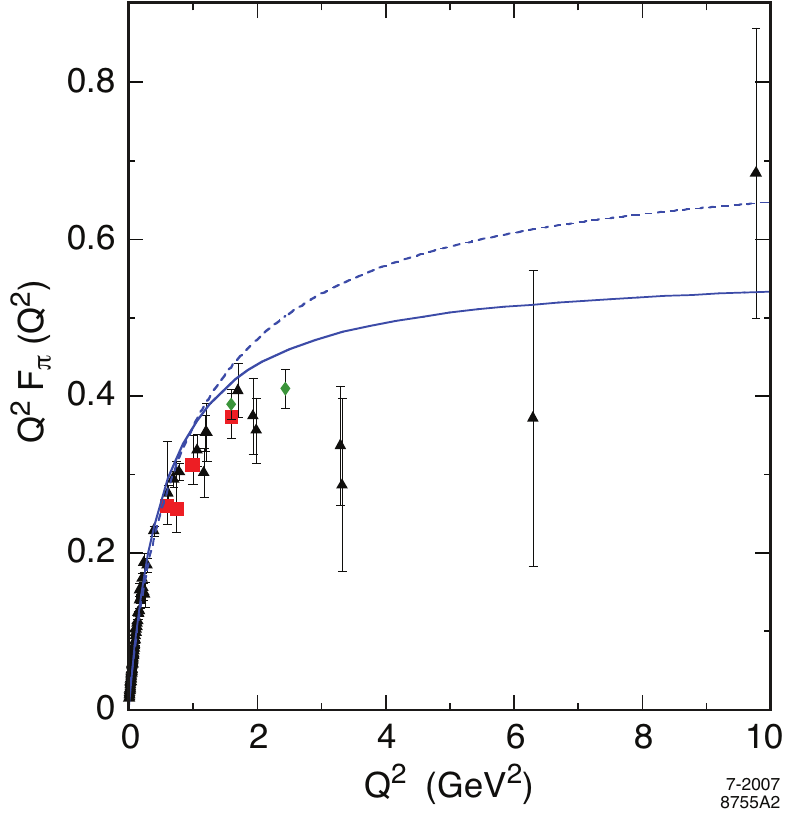}
\caption{Space-like scaling behavior for $Q^2 F_\pi(Q^2)$ as a function of $Q^2 = -q^2$.
The continuous line is the prediction of the soft-wall model for  $\kappa = 0.375$ GeV.
The dashed line is the prediction of the hard-wall model for $\Lambda_{\rm QCD} = 0.22$ GeV. 
The triangles are the data compilation  from Baldini  {\it et al.}~\cite{Baldini:1998qn},  the  boxes  are JLAB 1 data~\cite{Tadevosyan:2007yd} and  diamonds are JLAB 2
 data~\cite{Horn:2006tm}.}
\label{fig:PionQ2FFSL}
\end{figure}
Since the pion mode couples to a twist-two boundary
interpolating operator which creates a  two-component hadronic bound state, the
form factor is given in the SW model by the simple monopole form (\ref{eq:monopole}) corresponding
to $n=2$
\begin{equation} \label{eq:pionFFSW}
 F_\pi(Q^2) = \frac{4 \kappa^2}{4 \kappa^2 + Q^2},
 \end{equation}
 the well known vector dominance model with the leading $\rho$ resonance.

\begin{figure}[ h]
\centering
\includegraphics[angle=0,width=8.4cm]{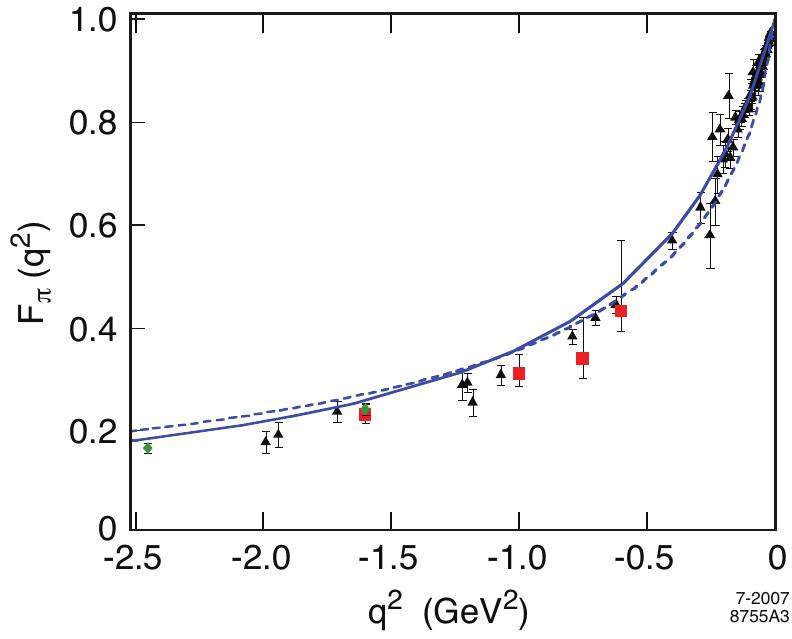}
\caption{Space-like behavior of the pion form factor $F_\pi(q^2)$ as a function of $q^2$ for $\kappa = 0.375$ GeV  and  $\Lambda_{\rm QCD} = 0.22$ GeV. Continuous line: soft-wall model, dashed line: hard-wall model. Triangles are the data compilation  from Baldini  {\it et al.}~\cite{Baldini:1998qn},  boxes  are JLAB 1~\cite{Tadevosyan:2007yd} and diamonds are JLAB 2~\cite{Horn:2006tm}.}
\label{fig:PionFFSL}
\end{figure}

\begin{figure}[h]
\centering
\includegraphics[angle=0,width=8.4cm]{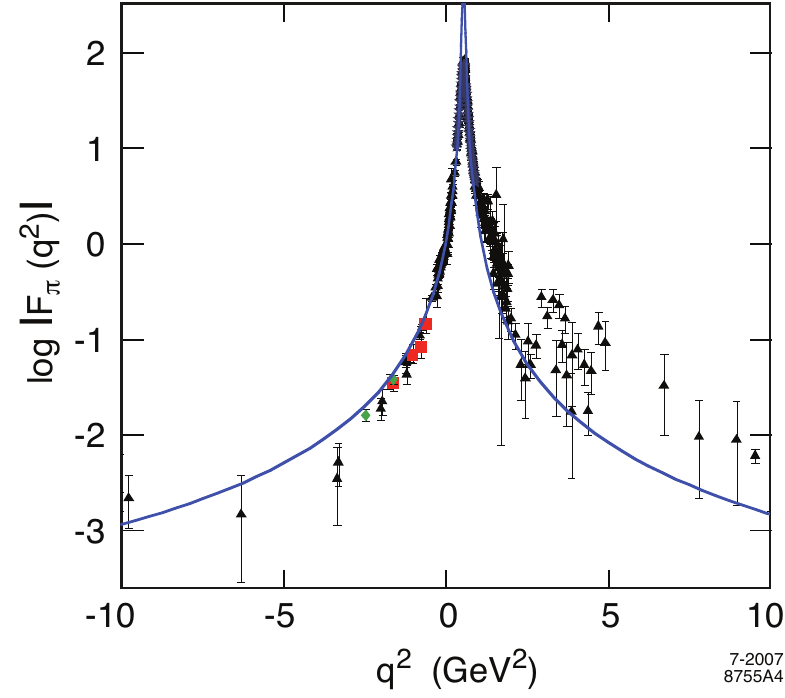}
\caption{Space and time-like behavior of the pion form factor $\log\left(\vert F_\pi(q^2)\vert\right)$ as a function of $q^2$ for $\kappa = 0.375$ GeV in the  soft-wall model. Triangles are the data compilation  from Baldini  {\it et al.}~\cite{Baldini:1998qn},  boxes are JLAB 1~\cite{Tadevosyan:2007yd} and  diamonds are JLAB 2~\cite{Horn:2006tm}.}
\label{fig:PionFFLog}
\end{figure}

The hadronic scale is evaluated
 by fitting the space-like data for the form factor as shown in Figure \ref{fig:PionQ2FFSL}, where we plot
 the product $Q^2 F_\pi(Q^2)$ for the soft and hard-wall holographic models.  Both models would seem to describe the overall behavior of the space-like data, with a better high-$Q^2$ description of
 scaling with the SW model. However, when  the low energy data is examined in detail the SW model gives a noticeable better description as shown in Figure \ref{fig:PionFFSL}.
When the results for the pion form factor are analytically continued to the time-like region, $q^2 \to -q^2$ we obtain the results shown in Figure \ref{fig:PionFFLog} for $\log\left(\vert F_\pi(q^2)\vert\right)$. The monopole form of the SW model reproduces  well the $\rho$ peak  with $M_\rho = 2 \kappa = 750$ MeV and the
scaling behavior in the space-like region, but it does not give rise to the
additional structure found in the time-like region, since the simple dipole form (\ref{eq:pionFFSW})
is saturated by the $\rho$ pole. The unconfined propagator $J(Q^2, z) = z Q K_1(z Q)$
used in our description of the hard wall model do not give rise to a pole structure in the time like region, and thus fails to reproduce the $\rho$ structure. 
When a small anomalous dimension is introduced,
$\tau = 2 \pm \epsilon$, the formula (\ref{eq:FFSWtau}) for the form factor should be used instead. For a small deviation from the canonical value of two, the behavior in the space like region remains almost identical, but the form-factor in the time like region oscillates with infinite amplitude, and thus gives an unphysical  solution. 

\begin{figure}[h]
\centering
\includegraphics[angle=0,width=8.0cm]{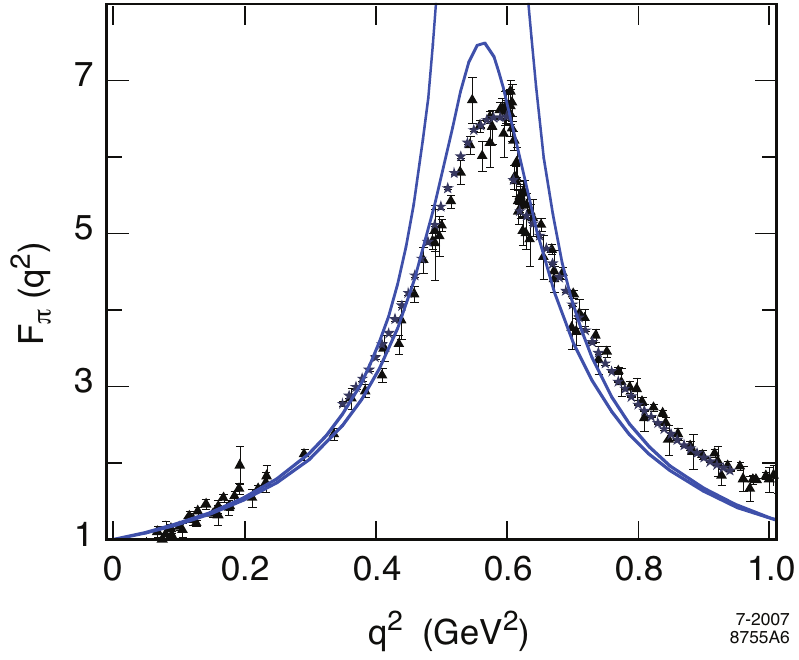}
\caption{Time-like pion form factor $\vert F_\pi(q^2)\vert$ near the $\rho$ peak as a function of $q^2$ for $\kappa = 0.375$ GeV in the soft-wall model. The upper curve correspond to $\Gamma = 0$ and the lower to $\Gamma = 100$ MeV. The  data is from the compilation of
Baldini {\it et al.}~\cite{Baldini:1998qn}.}
\label{fig:PionFFGamma}
\end{figure}
\begin{figure}[ h]
\centering
\includegraphics[angle=0,width=7.0cm]{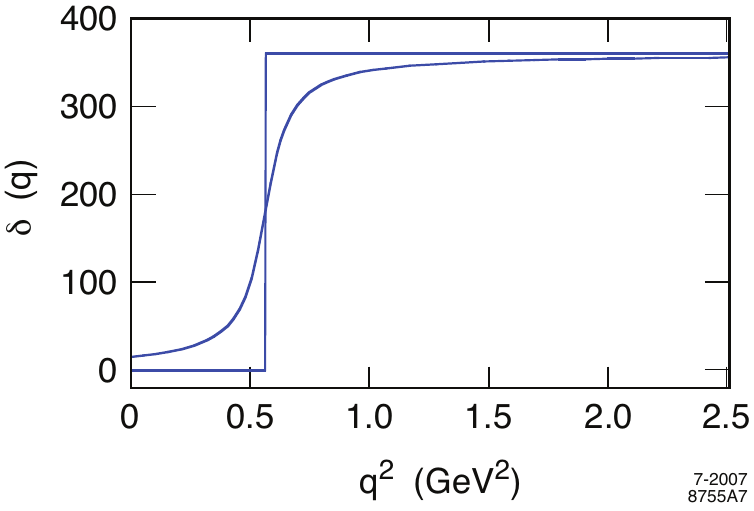}
\caption{Time-like pion form factor phase $\delta_\pi(q)$ as a function of $q^2$ for $\kappa = 0.375$ GeV in the soft-wall model. The sharp curve corresponds to $\Gamma = 0$ and the smooth curve to $\Gamma = 100$ MeV. }
\label{fig:PionFFphase}
\end{figure}

In the strongly coupled semiclassical gauge/gravity limit hadrons have zero widths  and are stable.
One can nonetheless modify the formula (\ref{eq:pionFFSW}) to introduce a finite width $\Gamma$
\begin{equation} 
 F_\pi(Q^2) = \frac{4 \kappa^2}{4 \kappa^2 - q^2 - 2 i \kappa \Gamma},
 \end{equation}
 to compare with the pion form factor data near the $\rho$ peak. The result is shown in 
 Figure \ref{fig:PionFFGamma}. The corresponding phases for the pion form factor are shown in Figure
  \ref{fig:PionFFphase}.

Using our results for the pion form factor one can extract the value of the mean pion charge
radius squared through the small $Q^2$ expansion
\begin{equation}
F_\pi(Q^2) =  1 - \frac{1}{6} \langle r_\pi^2 \rangle \, Q^2 + \mathcal{O}(Q^4),
\end{equation}
and thus
\begin{equation}
\langle r_\pi^2 \rangle = - 6 \frac{d F_\pi(Q^2)}{d Q^2} \Big\vert_{Q^2 = 0}.
\end{equation}
For the soft wall model the form factor is given by the monopole form (\ref{eq:pionFFSW}) 
and we find
\begin{equation}
\langle r_\pi^2 \rangle_{SW} = \frac{3}{2 \kappa^2} \simeq 0.42 ~ {\rm fm}^2,
\end{equation}
compared with the PDG value $\langle r_\pi^2 \rangle = 0.45(1) ~ {\rm fm}^2$.~\cite{Eidelman:2004wy}
In the hard-wall model  discussed in this paper the electromagnetic current is not confined
and propagates in conformal metrics according to (\ref{eq:AdSJ}). To compute 
the second moment
of the charge distribution $\langle r_\pi^2 \rangle$ in this case, we expand $J(Q^2, z)$ in (\ref{eq:J})  for small values of $Q^2$ 
\begin{equation}
J(Q^2,z) = 1 + \frac{z^2 Q^2}{4} 
\left[ 2 \gamma  - 1 + \ln\left(\frac{z^2 Q^2}{4}\right) \right]+ \mathcal{O}^4,
\end{equation}
where $\gamma$, the Euler-Mascheroni constant, has the value $0.5772\dots$  As there is no scale
present in the expression for the bulk-to-boundary propagator (\ref{eq:J}),   the value of $\langle r_\pi^2 \rangle$ diverges logarithmically 
and thus the slope of the form factor is logarithmically divergent, although the predicted shape is consistent with the data except for very low $Q^2$. This formal problem in defining $\langle r_\pi^2\rangle$ does not appear if one uses Neumann boundary conditions for the HW model, and 
$\langle r_\pi^2 \rangle  \sim 1/\Lambda_{\rm QCD}^2$.

\section{The Pion Decay Constant}

The pion decay constant is given in terms of the matrix element of the axial isospin
current $J^{\mu 5 a}$ between a physical pion and the vacuum state~\cite{Peskin:1995ev}
\begin{equation}
\left\langle 0 \left\vert J^{\mu 5}_a(x) \right\vert \pi_b(P)\right\rangle =
- i P^\mu f_\pi \delta_{ab} e^{-i P \cdot x},
\end{equation}
Axial and vector currents are 
\begin{equation}
J^{\mu 5 a} = \bar Q \gamma^\mu \gamma^5 t^a Q,
~~~ J^{\mu a} = \bar Q \gamma^\mu t^a Q
\end{equation}
with  $Q$ the quark doublet
\begin{equation} Q =
\begin{pmatrix}
 \psi_u \\ \psi_d
\end{pmatrix},
\end{equation}
and $t^a = \tau^a/2$ the generators of isospin rotations.

Consider the amplitude for the weak decay of a charged pion
$\bigl\langle 0 \bigl\vert J_W^+(0) \bigr\vert \pi^-(P^+, \vec P_\perp) \bigr\rangle$,
where $J_W^+$ is the flavor changing weak current
\begin{equation} \label{eq:JW}
J_W^+ = \bar\psi_u \gamma^+ \tfrac{1}{2} (1 - \gamma_5) \psi_d.
\end{equation}
It follows that~\cite{ft:Peskin}
\begin{equation} \label{eq:Jud}
\left\langle 0 \left\vert\bar\psi_u \gamma^+\tfrac{1}{2}(1-\gamma_5)\psi_d 
\right\vert \pi^-\right\rangle
= i \frac{P^+ f_\pi}{\sqrt{2}}.
\end{equation}
The expression for the current operator (\ref{eq:JW}) in the particle number
representation follows from the expansion of the field operator
(\ref{eq:psiop}) in terms of quark creation and annihilation operators. 
The only terms which contribute to the annihilation process are the 
terms containing the product of
operators $d$ and $b$ with opposite spin.
Using the results listed in Appendix A for the quark spinor projections,
we obtain the effective current operator
\begin{equation}
J_W^+
= \int \frac{d q^+   d^2\vec q_\perp}{(2 \pi)^3}
  \int \frac{d q'^+ d^2\vec q~'_\perp}{(2 \pi)^3}
 ~d_{u\uparrow}(q)  b_{d\downarrow}(q'),
\end{equation}
in the limit of massless quarks.
The flavor changing operator $J_W^+$ annihilates a $d$-quark with momentum $q'$ and  
spin-down along the $z$ direction
and a $u$-antiquark momentum $q$ and spin-up.
Only the valence state
\begin{equation}
\bigl\vert d \bar u\bigr\rangle = \frac{1}{ \sqrt{N_C}} \frac{1}{\sqrt{2}}
 \sum_{c=1}^{N_C}
\left(b_{c\,d\downarrow}^\dagger d_{c\,u\uparrow}^\dagger
- b_{c\,d\uparrow}^\dagger d_{c\,u\downarrow}^\dagger\right) 
\bigl\vert 0 \bigr\rangle,
\end{equation}
with $L_z = 0$, $S_z = 0$, contributes to the decay of the $\pi^\pm$. 
Expanding the  hadronic initial state in the amplitude (\ref{eq:Jud})
into its Fock components we find
 \begin{equation} 
f_\pi =  2 \sqrt{N_C} \int_0^1 dx \int \frac{d^2 \mbf{k}_\perp}{16 \pi^3} 
~\psi_{\bar q q/\pi}(x, \mbf{k}_\perp),
\end{equation}
the light-cone equation which allows the exact computation of the pion decay
constant in terms of the valence pion light-front wave function~\cite{Lepage:1980fj}.

In terms of the distribution amplitude $\phi(x,Q)$
\begin{equation}
\phi(x,Q) = \int^{Q^2} \frac{d^2 \mbf{k}_\perp}{16 \pi^3}~\psi(x, \mbf{k}_\perp),
\end{equation}
it follows that
\begin{equation} \label{eq:phix} 
\phi_\pi(x, Q\to\infty)
= \frac{4}{\sqrt{3}\pi}  f_\pi \sqrt{x(1-x)}, 
\end{equation}
with
\begin{equation}
f_\pi = \frac{1}{8} \sqrt{\frac{3}{2}} \, R^{3/2} \lim_{\zeta \to 0}
\frac{\Phi(\zeta)}{\zeta^2}.
\end{equation}
since $\phi(x, Q \to \infty) \to 
\widetilde \psi(x,\mbf{b}_\perp \to 0)/\sqrt{4 \pi}$ as $\zeta\to 0$.
The pion decay constant depends only on the behavior of the AdS
string mode near the asymptotic boundary, $\zeta = z = 0$, and the
mode normalization. For the hard-wall truncated-space pion mode 
(\ref{eq:PhipiHW}) we find
\begin{equation}
f_\pi^{HW} = \frac{\sqrt{3}}{8 J_1(\beta_{0,k})} \, \Lambda_{\rm QCD}
= 91.7 ~{\rm MeV},
\end{equation}
for $\Lambda_{\rm QCD} = 0.22$ MeV. The corresponding result for the
soft-wall pion mode (\ref{eq:PhipiSW}) is
\begin{equation}
f_\pi^{SW} =  \frac{\sqrt{3}}{8} \, \kappa = 81.2 ~{\rm MeV},
\end{equation}
for $\kappa = 0.375$ MeV. The values of $\Lambda_{QCD}$ and $\kappa$ are
determined from the spacelike form-factor data as discussed above.
The experimental result for $f_\pi$ is extracted from the rate 
of weak $\pi$ decay and has the value 
$f_\pi = 92.4$ MeV~\cite{Eidelman:2004wy}. 

It is interesting to note that the pion distribution amplitude
predicted by AdS/QCD (\ref{eq:phix})
has a quite different $x$-behavior than the
asymptotic distribution amplitude predicted from the PQCD
evolution~\cite{Lepage:1979zb} of the pion distribution amplitude
$\phi_\pi(x,Q \to \infty)= \sqrt 3  f_\pi x(1-x) $.   This observation appears to be consistent with the results of the Fermilab diffractive dijet experiment~\cite{Aitala:2000hb} which shows a broader  $x$ distribution for the dijets at small transverse momentum $k_\perp \le 1 $ GeV.  The broader
shape of the pion distribution increases the magnitude of the
leading twist perturbative QCD prediction for the pion form factor
by a factor of $16/9$ compared to the prediction based on the
asymptotic form, bringing the PQCD prediction  close to the
empirical pion form factor~\cite{Choi:2006ha}.

\section{Conclusions}

One of the key difficulties in studies of quantum chromodynamics has been the absence of an analytic first approximation to the theory which not only can  reproduce the hadronic spectrum, but also provides a good description of hadron wavefunctions.  The AdS/CFT  correspondence provides an elegant semi-classical approximation to QCD, which incorporates both color confinement and the conformal short-distance behavior appropriate for a theory with an infrared fixed point.   Since the hadronic solutions are controlled by their twist dimension $z^\tau$ at small $z$, one  also reproduces dimensional counting rules for hard exclusive processes. 
The AdS/CFT approach thus allows one to construct a model of hadrons
which has both confinement at large distances and the conformal
scaling properties which reproduce dimensional counting rules for
hard exclusive reactions.  The fundamental equations of AdS/CFT for mesons have
the appearance of a radial Schr\"odinger Coulomb equation, but they are
relativistic, covariant, and analytically tractable. The eigenvalues of the AdS/CFT equations provide a  good description of the meson and baryon
spectra for light quarks~\cite{deTeramond:2005su, Brodsky:2006uq, deTeramond:2006xb}, and its eigensolutions provide a remarkably simple but realistic model of their valence
wavefunctions.   One can also derive analogous equations for baryons composed of massless quarks using a Dirac matrix representation for the baryon system~\cite{Brodsky:2007pt}.

The lowest stable
state of the AdS equations is determined by the Breitenlohner-Freedman bound~\cite{Breitenlohner:1982jf}.  We can model confinement  by imposing Dirichlet boundary
conditions at $\phi(z = 1/\Lambda_{\rm QCD}) = 0.$ The eigenvalues are then given in terms of the roots of the Bessel functions:
$\mathcal{M}_{L,k} = \beta_{L,k} \Lambda_{\rm QCD}$.   Alternatively, one can  introduce a dilaton 
field~\cite{Karch:2006pv} which provides a confinement potential $-\kappa^2 \zeta^2$ to the effective
potential $V(\zeta).$  The resulting hadron spectra are given by linear Regge trajectories in the square of the hadron masses, $\mathcal{M}^2$, characteristic of the Nambu string model.  
The AdS/CFT equations are integrable and thus the radial and orbital excitations can be obtained from ladder operators~\cite{Brodsky:2007pt}.

In this work we have shown that the eigensolutions  $\Phi_H(z)$  of the AdS/CFT equations in the fifth dimension $z$ have a remarkable mapping to the light-front wavefunctions 
$\psi_H(x_i,\mbf{b}_{\perp i})$, the hadronic amplitudes which describe the valence constituents of hadrons in physical space time, but at fixed light-cone time $\tau=t+z/c=0$.   Similarly,  the AdS/CFT equations for hadrons can be mapped to equivalent light-front equations. 
The correspondence of AdS/CFT amplitudes to the QCD wavefunctions in light-front coordinates in physical space-time, provides an exact holographic mapping at all energy scales between string modes in AdS space and hadronic boundary states.
Most important, the eigensolutions of the AdS/CFT equation can be mapped to light-front equations of the hadrons in physical space-time, thus
providing an elegant description of the light hadrons at the amplitude level. 

The mapping of AdS/CFT string modes to light-front wave functions thus provides a remarkable analytic  first approximation to QCD.
Since they are complete and orthonormal, the AdS/CFT model
wavefunctions can also be used as a basis for the diagonalization of
the full light-front QCD Hamiltonian, thus systematically improving
the AdS/CFT approximation. 

We have also shown the correspondence between the expressions for current matrix elements in AdS/CFT with the corresponding expressions for form factors as given in the light-front formalism. In first approximation, where one takes the current propagating in a non-confining background, one obtains the Drell-Yan West formula for valence Fock states, corresponding to the interaction picture of the light-front theory.  Hadron form factors can thus be directly predicted from the overlap integrals in AdS space or equivalently by using
the Drell-Yan-West formula in physical space-time.   The form factor at high $Q^2$ receives its main contributions from small $\zeta \sim {1/ Q}$,
corresponding to small $\vert\mbf{b}_\perp\vert = {\cal O}(1/Q)$ and  $1-x  = {\cal O}(1/Q)$.
We have also shown how to improve this approximation  by studying the propagation of non-normalizable solutions representing the electromagnetic  current in a modified AdS 
confining metric, or equivalently in a dilaton background.   This improvement in the description of the current corresponds  in the light-front to multiple hadronic Fock states.  The introduction of the confined current implies that the timelike form factors of  hadrons will be mediated by vector mesons, including radial excitations.  The wavefunction of the normalizable vector meson states $\mathcal{A}(z)$ appearing in the spectral
decomposition of the Green's function  described in Appendix   C, which is dual to the non-normalizable photon propagation mode in AdS, is twist-3. This is the expected result for even parity axial mesons in QCD, or $L=1$ odd parity vector mesons composed of a scalar squark and anti-squarks.  In the case of quark-antiquark states, one also expects to find  $C= - 1$, twist-2 meson solutions for the zero helicity component of the $\rho$ with $S=1$ and $L=0$, which is supposed to give a dominant contribution to the $\rho$ form factor. 

We have applied our formulation to both the spacelike and timelike pion form factor.  The description of the pion form factor in the spacelike domain is in good agreement with experiment in both confinement models, the hard and the soft wall holographic models, but the hard wall model discussed here fails to reproduce the very low $Q^2$ data. In particular, the mean pion charge radius squared 
$\langle r_\pi^2\rangle$ diverges logarithmically.
In the soft wall model the time-like pion form factor exhibits a pole at the $\rho$ mass with zero width since hadrons are stable in this theory, but it does not reproduce the additional resonant structure
found in the time-like region. If one allows a width, the height of the $\rho$ pole is in reasonable agreement with experiment. 
The space-like Dirac form factor for the proton is also very well reproduced by the double-pole analytic expression shown in Appendix D for the case $N=3$. This will be discussed elsewhere.

The deeply virtual Compton amplitude in the  handbag approximation can be expressed as overlap of light-front wavefunctions~\cite{Brodsky:2000xy}.
The deeply virtual Compton amplitudes can be Fourier transformed to $b_\perp$ and $\sigma = x^-P^+/2$ space providing new
insights into QCD distributions~\cite{Burkardt:2005td,Ji:2003ak,Ralston:2001xs,Brodsky:2006in,Hoyer:2006xg}. The distributions in the light-front direction $\sigma$
typically display diffraction patterns arising from the interference of the initial and final state LFWFs
~\cite{Brodsky:2006in,Brodsky:2006ku}. 
All of these processes  can provide detailed tests of  the AdS/CFT LFWFs predictions.

The phenomenology of the AdS/CFT model is just begining, but it can
be anticipated that it will have many applications to QCD phenomena at the amplitude level.
For example, the model LFWFs  obtained from AdS/QCD provide a basis for understanding
hadron structure functions and fragmentation functions at the
amplitude level; the same wavefunctions can describe hadron
formation from the coalescence of co-moving quarks.  The spin
correlations which underly single and double spin correlations are
also described by the AdS/CFT eigensolutions.  The AdS/CFT hadronic
wavefunctions also provide predictions for the generalized parton distributions of hadrons
and their weak decay amplitudes from first principles.  The
amplitudes relevant to diffractive reactions could also be computed. We
also anticipate that the extension of the AdS/CFT formalism to heavy
quarks will allow a great variety of heavy hadron phenomena to be
analyzed from first principles.

\begin{acknowledgments}

This research was supported by the Department
of Energy contract DE--AC02--76SF00515. The research of GdT is supported in part
by an Aportes grant from Florida Ice \& Farm. We thank Oleg Andreev, Garth Huber, Cheng-Ryong Ji,
Leonid Glozman, Bernard Pire,  Anatoly Radyushkin, Giovanni Salm\`e and James Vary for helpful comments.

\end{acknowledgments}

\vspace{15pt}

\centerline{\bf \small Note Added}

\vspace{10pt}

After finishing this paper, two new papers discussing the pion form factor
including chiral symmetry breaking effects using the framework introduced in~\cite{Erlich:2005qh, DaRold:2005zs}  in AdS/CFT, were posted into arXiv. See~\cite{Kwee:2007dd, Grigoryan:2007wn}.  
In the approach of
~\cite{Erlich:2005qh, DaRold:2005zs} the axial and vector currents become the primary
entities as in effective chiral theory, in contrast with the framework described here where all mesons are quark-antiquark composites.

\appendix

\section{ Light-Cone Spinors}

In the light-cone (LC) formalism,  spinors are
written as~\cite{Lepage:1980fj}
\begin{eqnarray}
u^\uparrow(p) &=& \frac{1}{\sqrt{p^+}} 
\left(p^+ + \beta m + \vec\alpha_\perp \cdot \mbf{p}_\perp \right) \chi^\uparrow , \\ 
v^\uparrow(p) &=& \frac{1}{\sqrt{p^+}}
\left(p^+ - \beta m + \vec\alpha_\perp \cdot \mbf{p}_\perp \right) \chi^\downarrow,
\end{eqnarray} 
with representation matrices
\begin{equation}
\vec \alpha =
  \begin{pmatrix}
  0&   \vec\sigma\\
  \vec \sigma& 0
  \end{pmatrix},~~~~~
\beta =  
  \begin{pmatrix}
  I&   ~0\\
  0&   -I
  \end{pmatrix},~~~~~
\gamma_5 =
  \begin{pmatrix}
  0&   I\\
  I&   0
  \end{pmatrix},
\end{equation}
and spinors
\begin{equation}
\chi^\uparrow = \frac{1}{\sqrt{2}}
  \begin{pmatrix}
  1 \\ 0 \\1 \\0
  \end{pmatrix},~~~~~~~
\chi^\downarrow = \frac{1}{\sqrt{2}}
  \begin{pmatrix}
  0 \\ 1 \\0 \\-1
  \end{pmatrix}.
\end{equation}
The corresponding solutions for LC spinors $u^\downarrow(p)$ and
$v^\downarrow(p)$ 
follow from reversing the spin component. 
Useful Dirac matrix elements for the helicity spinors are~\cite{Lepage:1980fj}
\begin{eqnarray} \label{eq:sn1}
\bar u^\uparrow(\ell) \gamma^+ u^\uparrow(k)
&=& \bar u^\downarrow(\ell) \gamma^+ u^\downarrow(k) = 2 \sqrt{\ell^+ k^+} , \\
 \bar u^\downarrow(\ell) \gamma^+ u^\uparrow(k)
&=& \bar u^\uparrow(\ell) \gamma^+ u^\downarrow(k) = 0,
\end{eqnarray}
with $ \bar v_\alpha(\ell) \gamma^\mu v_\beta(k) = \bar u_\beta(k)  \gamma^\mu u_\alpha(\ell) $ and
\begin{eqnarray} \label{eq:sn2}
\bar v^\downarrow(\ell) \gamma^+ u^\uparrow(k)
&=& \bar v^\uparrow(\ell) \gamma^+ u^\downarrow(k) = 2 \sqrt{\ell^+ k^+}  , \\
 \bar v^\uparrow(\ell) \gamma^+ u^\uparrow(k)
&=& \bar v^\downarrow(\ell) \gamma^+ u^\downarrow(k) = 0.
\end{eqnarray}

It is simple to verify that
\begin{equation}
\tfrac{1}{2}(1-\gamma_5) \chi^\uparrow = 0, ~~~
\tfrac{1}{2}(1-\gamma_5) \chi^\downarrow = \chi^\downarrow.
\end{equation}
Thus the chiral projection of the boosted LC spinors $u(p)$ and $v(p)$
in the limit of massless quarks 
\begin{eqnarray} \label{eq:u}
\tfrac{1}{2} (1-\gamma_5)u^\downarrow (p) = u^\downarrow (p), ~~~
\tfrac{1}{2} (1-\gamma_5)u^\uparrow (p) = 0, \\ \label{eq:v} \negthinspace
\tfrac{1}{2} (1-\gamma_5)v^\downarrow (p) = 0, ~~~~
\tfrac{1}{2} (1-\gamma_5)v^\uparrow(p) = v^\uparrow (p),
\end{eqnarray}
which follows from the fact that $\gamma_5$ commutes with $\vec\alpha_\perp$, but
not with $\beta$. Consequently, in the
massless limit we find that LC quark spinors with positive spin 
projection along the $z$-component also transform as right-handed
spinors and LC spinors with negative spin projection are left-handed
spinors. Also spinors representing massless antiquarks with positive 
spin component along the $z$-direction are left-handed and
right-handed if polarized along the negative $z$-direction.

\section{A Useful Integral}

Consider the integral
\begin{equation} \label{eq:intJz}
J(Q^2, z) =\int_0^1 dx \, J_0\negthinspace\left(z Q
\sqrt{\frac{1-x}{x}}\right) 
\end{equation}
Changing the variable $x$ according to
$x = \frac{Q^2}{t^2 + Q^2}$,
we recast (\ref{eq:intJz}) as
\begin{equation}
J(Q^2, z) = 2 Q^2 \int_0^\infty \frac{t J_0(z t)}
 {\left(t^2 + Q^2\right)^{2}} \, dt,
 \end{equation}
 which is a Hankel-Nicholson type integral \cite{AS:p488},
\begin{equation}
\int_0^\infty \frac{t^{\nu+1} J_\nu(z t)}
 {\left(t^2 + a^2\right)^{\mu+1}} \, dt  =
\frac{z^\mu a^{\nu - \mu}}{2^\mu \Gamma(\mu+1)} K_{\nu-\mu}(a z).
\end{equation}
Thus
\begin{equation}
J(Q^2, z) = Q z K_1(z Q).
\end{equation}

\section{Bulk-to-Boundary Propagator in the Soft Wall Model}

The bulk-to-boundary propagator in the soft-wall holographic model 
follows from the solution of the wave equation
\begin{equation} \label{eq:AdSJkappaA}
\left[z^2 \partial_z^2   -  \left( 1 + 2 \kappa ^2 z^2   \right) z \partial_z - Q^2 z^2 \right] \!
J_\kappa(Q^2, z) = 0,
\end{equation}
describing
the propagation of an electromagnetic probe in AdS space coupled to a dilaton background field
$\varphi(z) = \kappa^2 z^2$.
The solution to (\ref{eq:AdSJkappaA})
subject to the boundary conditions
\begin{equation} 
J_\kappa(Q^2 = 0, z) = J_\kappa(Q^2, z = 0) = 1,
\end{equation}
is
 \begin{equation} 
J_\kappa(Q^2, z) = \Gamma \! \left(1+ \frac{Q^2}{4 \kappa^2}\right) 
U \! \left(\frac{Q^2}{4 \kappa^2}, 0 , \kappa^2 z^2\right).
\end{equation}

Using the integral representation of the confluent hypergeometric function 
$U(a,b,z)$~\cite{AS:p505}
 \begin{equation}
 \Gamma(a)  U(a,b,z) =  \int_0^\infty \! e^{- z t} t^{a-1} (1+t)^{b-a-1} dt,
  \end{equation} we find
  \begin{equation} \label{eq:IntRepJt}
  J_\kappa(Q^2, z) = \frac{Q^2}{4 \kappa^2}
  \int_0^\infty \! e^{- \kappa^2 z^2 t} \left(\frac{1+t}{t}\right)^{-\frac{Q^2}{4 \kappa^2}}
  \frac{dt}{t(1+t)}.
  \end{equation}
  Making the change of variable $t = \frac{Q^2}{4 \kappa^2} \,\mu$ and
 taking the limit  $Q^2 \gg \kappa^2$  
 \begin{equation}
 J_\kappa(Q^2 \to \infty ,z) =
 \int_0^\infty  \frac{d\mu}{\mu^2} \, \exp\left(- \frac{z^2 Q^2}{4} \mu - \frac{1}{\mu}\right).
 \end{equation}
 Comparing the above expression with the integral representation of the modified Bessel function
  $K_\nu(z)$
 \begin{equation}
 K_\nu(z) = \frac{1}{2} \left(\frac{z}{2}\right)^\nu
\int_0^\infty  \frac{e^{-t - \frac{z^2}{4 t}}}{t^{\nu+1} }\, dt ,
 \end{equation}
 we find 
 \begin{equation}
 \lim_{Q^2 \to \infty}J_\kappa(Q^2 , z) = z Q K_1(z Q).
 \end{equation}
In the large $Q^2$ limit the electromagnetic probe in AdS space decouples from the dilaton background.

\subsection{Green's Function Formalism}

In a finite spatial dimension the Green's function has an spectral decomposition in
terms of eigenfunctions. Normalizable spin-one  modes propagating on AdS$_5$
have polarization and plane waves along Minkowski coordinates 
$\mathcal{A}(x,z)_\mu = e^{- i P \cdot x} \mathcal{A}(z) \epsilon_\mu$, with
invariant mass $P_\mu P^\mu = \mathcal{M}^2$. The eigenfunctions are
determined by the eigenvalue equation 
\begin{equation} \label{eq:AdSASW}
\left[- z^2 \partial_z^2   +  \! \left( 1 + 2 \kappa ^2 z^2  \right) \! z \partial_z \right] \! \mathcal{A}_n(z) 
=  M_n^2 \, z^2  \mathcal{A}_n(z),
\end{equation}
with solutions
\begin{equation}
\mathcal{A}_n(z) = \frac{\kappa^2}{R^{1/2}} 
\sqrt{\frac{2}{n\!+\!1}} \, z^2 L^1_n\left(\kappa^2 z^2\right),
\end{equation}
and eigenvalues
\begin{equation} \label{eq:Mn}
\mathcal{M}_n^2 = 4 \kappa^2(n+1).
\end{equation}
The eigenfunctions $\mathcal{A}_n$ are orthonormal
\begin{equation}
R \! \int \! \frac{dz}{z} \, e^{- \kappa^2 z^2}  \mathcal{A}_n(z) \mathcal{A}_m(z) = \delta_{n m},
\end{equation}
and satisfy the completeness relation
\begin{equation} \label{eq:crSW}
R \sum_{n} \frac{e^{- \kappa^2 z'^2}} {z'} A_n(z') A_n(z) = \delta\left(z - z'\right).
\end{equation}

The Green's function is a solution of the homogeneous equation
\begin{multline} \label{eq:AdSGSW}
\left[-z^2 \partial_z^2 +  \left( 1 + 2 \kappa ^2 z^2   \right) \! z  \, \partial_z - q^2 z^2 \right] 
G(z,z';q) =  \\ - \delta\!\left(z-z'\right) ,
\end{multline}
where $q^2 = - Q^2 \le 0$.
Since the eigenfunctions $\mathcal{A}_n$ form a complete basis we can perform an
spectral expansion of the Green's function
\begin{equation}
G(z,z';q) = R \sum_{n m} a_{n m}(Q) 
\frac{ \, e^{- \kappa^2 z'^2}}{z'}  \mathcal{A}_n(z') \mathcal{A}_m(z).
\end{equation} 
Substituting the above expression in  (\ref{eq:AdSGSW}) and using (\ref{eq:crSW}) the
expansion coefficients $a_{nm}$ are determined.  We find
\begin{equation} \label{eq:GFSW}
G(z,z';q) = \sum_n \frac{\mathcal{A}_n(z') \mathcal{A}_n(z)}{q^2 - M_n^2 + i \epsilon}.
\end{equation}

\subsection{Relation with the Results of Grigoryan and Radyushkin}

Performing the change of variable $ t = \frac{x}{1-x}$  in the integral representation for the
bulk-to-boundary propagator (\ref{eq:IntRepJt}), and integrating by parts the resulting expression,
there follows
\begin{equation} \label{eq:IntRepJx}
J_\kappa(Q^2,z) = \kappa^2 z^2 \int_0^1 \! \frac{dx}{(1-x)^2} \, x^{\frac{Q^2}{4 \kappa^2}} 
e^{-\kappa^2 z^2 x/(1-x)},
\end{equation}
the result found by Grigoryan and Radyuskin in~\cite{Grigoryan:2007my}.  It was also  pointed out in~\cite{Grigoryan:2007my} that
the integrand in (\ref{eq:IntRepJx}) contains the generating function of the associated
Laguerre Polynomials
\begin{equation}
\frac{e^{-\kappa^2 z^2 x/(1-x)}}{{(1-x)^{k+1}}} = \sum_{n=0}^\infty 
L_n^k (\kappa^2 z^2) \, x^n,
\end{equation}
and thus $J_\kappa(Q^2, z)$ can be expressed as a sum of poles~\cite{Grigoryan:2007my}
\begin{equation}
J_\kappa(Q^2, z) = 4 \kappa^4 z^2 \sum_{n=0}^{\infty} 
\frac{L_n^1(\kappa^2 z^2)}{Q^2 + M_n^2},
\end{equation}
with $M_n^2 $ given by (\ref{eq:Mn}). The above result also follows immediately from
 (\ref{eq:GFSW}) using the expression
 \begin{equation}
 J(Q^2, z)  = J(Q^2, 0)  \lim_{z'\to 0} \frac {R}{z'} \, e^{-\kappa^2 z'^2}
 \partial_{z'} G(z,z';q),
 \end{equation}
 which defines the bulk-to-boundary propagator in terms of the
 Green's function~\cite{Hong:2004sa}.

\section{Analytic Solution of Structure Functions and Form Factors for Arbitrary Twist in the Soft Wall Model}

A string mode $\Phi_\tau$  in the soft-wall model  representing the lowest radial $n = 0$ node
\begin{equation} 
\Phi_\tau(z) = \frac{1}{R^{3/2}} \sqrt{\frac{2}{\Gamma(\tau \! - \! 1)} } \, \kappa^{\tau -1} z ^{\tau}.
\end{equation} 
 is normalized according to
\begin{equation} \label{eq:PhiNormtau}
\langle\Phi_\tau\vert\Phi_\tau\rangle = R^3 \int \frac{dz}{z^3} \, e^{- \kappa^2 z^2} \Phi_\tau(z)^2  = 1.
\end{equation}
Since the field $\Phi_\tau$ couples to a local hadronic interpolating operator of twist $\tau$ 
defined at the asymptotic boundary of AdS space,
the scaling dimension of $\Phi_\tau$ is $\tau$. 

\subsection{The Structure Function}

To compute the structure function $q(x)$ 
\begin{equation}
\int_0^1 \! dx \,q(x) = 1,
\end{equation}
corresponding to the the string mode $\Phi_\tau$, we substitute the the integral
representation of the unit operator
in the warped geometry of the soft-wall model 
\begin{equation} \label{eq:IntRep1}
\mbf{1} = \kappa^2 z^2 \int_0^1 \! \frac{dx}{(1-x)^2} \, 
e^{-\kappa^2 z^2 x/(1-x)},
\end{equation}
in  the normalization condition (\ref{eq:PhiNormtau}):  $\langle\Phi_\tau\vert\mbf{1}\vert\Phi_\tau\rangle$.
Upon integration over the coordinate $z$ we find
\begin{equation} \label{eq:sftau}
q(x) = (\tau \!-\!1) \, (1 - x)^{\tau-2},
\end{equation}
which gives a constant $x$-dependence for a two-parton hadronic bound state.
Equation (\ref{eq:IntRep1}) follows from (\ref{eq:IntRepJx}) in the limit $Q=0$.

\subsection{Hadronic Form Factor}

Likewise, we can compute the hadronic form factor in the soft wall holographic model 
\begin{equation} \label{eq:AdSFFtau}
F(Q^2) = R^3 \int \frac{dz}{z^3} \, e^{- \kappa^2 z^2} \Phi_\tau(z) J_\kappa(Q^2, z) \Phi_\tau(z) ,
\end{equation}
by substituting the integral representation  (\ref{eq:IntRepJx}) for
$J_\kappa(q,z)$ in (\ref{eq:AdSFFtau}) and integrating over the variable  $z$. We find the result
\begin{equation} \label{eq:FFxSW}
F(Q^2) =  \int_0^1 dx \, \rho(x,Q),
\end{equation}
where
\begin{equation} 
\rho(x,Q) = (\tau \!-\!1) \, (1 - x)^{\tau-2} \, x^{\frac{Q^2}{4 \kappa^2}} .
\end{equation}
The integral (\ref{eq:FFxSW}) can be expressed in terms of Gamma functions.
The final result for the form factor is
\begin{equation} \label{eq:FFSWtau}
F(Q^2) = \Gamma(\tau)  
\frac{\Gamma\left(1\! + \! \frac{Q^2}{4 \kappa^2}\right)}{\Gamma\left(\tau \! + \! \frac{Q^2}{4 \kappa^2}\right)}.
\end{equation}

In the absence of anomalous dimensions the twist is an integer, $\tau = N$,   and we can simplify 
(\ref{eq:FFSWtau}) by using the recurrence formula
\begin{equation} 
\Gamma(N+z) = (N - 1 + z) (N - 2 + z) \dots (1 + z) \Gamma(1+z).
\end{equation}
We find
\begin{eqnarray} \label{eq:monopole}
 F(Q^2) \! &=& \! \frac{1}{1 + \frac{Q^2}{4 \kappa^2}},  ~~~N = 2,  ~~~\\ \label{eq:dipole}
F(Q^2) \! &=& \!  \frac{2}{\left(1  + \! \frac{Q^2}{4 \kappa^2}\right) \!
 \left(2 + \! \frac{Q^2}{4 \kappa^2}\right)},
~~~ N =  3, \\  
  &   &  ~~~~~~~~   \cdots \nonumber \\  \label{eq:nminusonepole}
 F(Q^2) \!&=& \! \frac{(N-1)!}{\left(1 + \!\frac{Q^2}{4 \kappa^2} \right) \!
 \left(2 + \! \frac{Q^2}{4 \kappa^2}  \right)  \! \cdots \!
       \left(N \! - \! 1 \! + \! \frac{Q^2}{4 \kappa^2}  \right)},    N, ~~~~~~~
\end{eqnarray}
which is expressed as a $N-1$ product of poles, corresponding to the first $N-1$ states along the vector meson radial trajectory. 
For large $Q^2$ it follows that
\begin{equation}
F(Q^2) \to (N-1)! \left[\frac{4 \kappa^2}{Q^2}\right]^{(N-1)}, 
\end{equation}
and we recover the power law counting rules for hard scattering.

\section{Contributions to Meson Form Factors and Structure Functions at Large Momentum Transfer in AdS/QCD}

The  form factor of a hadron  at large $Q^2$ arises from the small $z$ kinematic domain in  AdS space. According to the AdS/CFT duality, this corresponds to small distances $x^\mu x_\mu \sim {1/Q^2}$ in physical space-time, the domain where the current matrix elements are controlled by the conformal
twist-dimension, $\Delta$,  of the hadron's interpolating operator.   In the case of the front form, where $x^+=0$,  this corresponds to small transverse separation $x^\mu x_\mu =  -\mbf{x}^2_\perp.$  

As we have shown in~\cite{Brodsky:2006uq}, one can use holography to map the functional form of the string modes  $\Phi(z)$ in  AdS space to the light front wavefunctions in physical  space time by identifying $z$ with the  transverse variable $ \zeta =\sqrt{x\over 1-x} |\vec \eta_\perp | .$  Here  $\vec \eta_\perp = \sum^{n-1}_{i=1}  x_i  \mbf{b}_{\perp i}$ is the weighted impact separation,  summed over the impact separation of the spectator constituents.  The leading large-$Q^2$  behavior of form factors in AdS/QCD arises from small $\zeta \sim 1/ Q$,  corresponding to small transverse separation. 

For the case of a meson with two constituents the form factor can be written in terms of an effective light-front transverse density in impact space. From (\ref{eq:FFeta})
\begin{equation} \label{eq:bxFF}
F(q^2) = \int^1_0 \! dx  \int  \!d b^2 \, \tilde \rho(x, b, Q),
\end{equation}
with $\tilde \rho(x, b, Q) = \pi J_0\left(b \,Q (1-x)\right) \vert \tilde \psi(x, b) \vert^2$ and
$b = \vert \mbf{b}_\perp \vert$.   The kinematics are  illustrated  in figure 7 for the case of a meson with two constituents in the soft-wall model
\begin{equation} 
\tilde\psi_{q \bar q/\pi}(x, \mbf{b}_\perp)  
= \frac{\kappa}{\sqrt{\pi}} \sqrt{x(1-x)}~e^{-\half \kappa^2 x(1-x) \mbf{b}_\perp^2},
\end{equation}
where the Gaussian form of the LFWF favors short distance configurations with small $\zeta^2  = b^2_\perp x(1-x) \sim 1/ Q^2$.
Since we are mainly interested in studying the contribution from different regions to  the form factor at large $Q^2$, we have replaced the modified bulk-to-boundary propagator $J_\kappa(Q, z)$
(\ref{eq:Jkappa}) by its large $Q^2$ form $z Q K_1(z Q)$.
One sees a shift of the integrand $\tilde  \rho(x, b, Q)$  toward small $\vert \mbf{b}_\perp \vert$ and small
$1-x$ at high $Q^2.$  A similar behavior is observed for the LFWF obtained from the hard wall model.

\begin{figure}[h] \label{fig:bxFF} 
\centering
\includegraphics[angle=0,width=8.4cm]{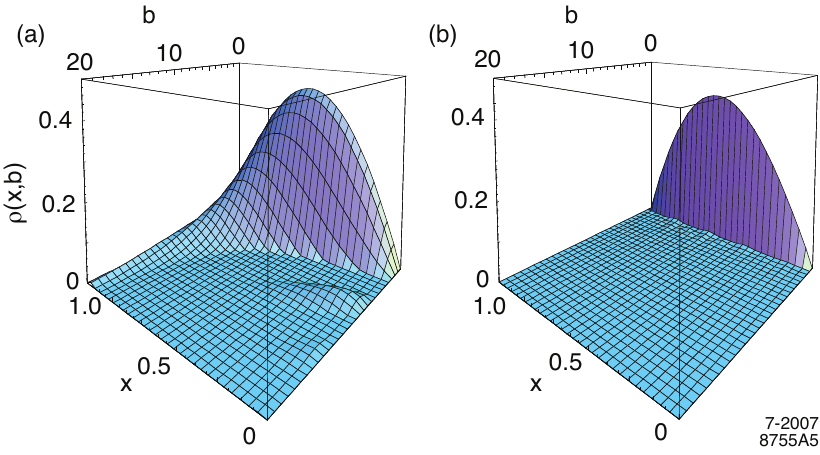}
\caption{Effective partonic density $ \tilde\rho(x,b,Q)$ in terms of the
 longitudinal momentum fraction $x$, the transverse relative impact
variable $b = \vert \mbf{b}_\perp\vert$ and momentum transfer $Q$ for the soft
wall model. As $Q$ increases the distribution becomes increasingly important near
$x=1$ and $\mbf{b}_\perp =0$. This is illustrated in (a) for $Q$ = 1 GeV/c.
At very large $Q$ (figure (b)), the distribution is peaked
towards $\mbf{b}_\perp = 0$. 
The value of $\kappa$ is 0.375 GeV.}
\end{figure}

The physical situation in momentum space  is somewhat more complicated.
The LFWF in $\mbf{k}_\perp$ space is the Fourier transform
\begin{equation}  
\psi(x, k) = 4 \pi^{3/2} \int_0^\infty \! b \, db \, J_0(k b) \, \tilde \psi(x,b),
\end{equation}
with $k = \vert \mbf{k}_\perp \vert$. Thus 
\begin{equation}  \label{eq:LFWFkSW}
\psi_{\bar q q/\pi}(x, \mbf{k}_\perp) = \frac{4 \pi}{\kappa \sqrt{x(1-x)}} 
~e^{- \frac{\mbf{k}_\perp^2}{2 \kappa^2 x(1-x)}} ,
\end{equation}
with exponential fall off at high relative transverse momentum.
The Drell-Yan-West light-front expression (\ref{eq:DYW}) for the form factor for a two-parton
bound state has the form 
\begin{multline} \label{eq:kFF}
F(q^2)  =  \int_0^1 dx \int \frac{d^2 \mbf{k}_\perp}{16 \pi^3} \\  \times
\sum_q e_q \psi^*_{P'} (x, \mbf{k}_\perp \! + (1 - x) \mbf{q}_\perp) \,
\psi_P (x, \mbf{k}_\perp).
\end{multline}
The form factor can be written in terms of a density in $x, \mbf{k}_\perp$ space:
\begin{equation} \label{eq:kxFF}
F(Q^2) = \int_ 0^1 \! dx \! \int  \! dk^2 \rho(x,k,Q),
\end{equation}
where
\begin{equation}
\rho(x,k,Q) =  \frac{1}{\kappa^2}
\frac{1}{x(1-x)}  \,
 I_0 \! \left(\frac{  k Q}{{\kappa^2}{x}}\right)
e^{- \frac{ (1-x) Q^2 }{2 \kappa^2 x}}
\, e^{- \frac{\mbf{k}_\perp^2}{ \kappa^2 x(1-x)}}.
 \end{equation}
In this case the large $Q^2$ behavior comes from the regime $1-x \sim \kappa^2/Q^2$ and $k^2_\perp \sim {x (1-x) \kappa^2}  \sim \kappa^4/Q^2.$  Note that the $x \to 1$ domain for finite $k_\perp$ (or nonzero quark mass)  implies $x_j = k^+_j/P^+ \to 0$ for each spectator, a regime of high relative longitudinal momentum since the limit $k^+ \to 0$ requires $k^z \simeq -k^0;$ i.e., $k^2_z \gg \mbf{k}^2_\perp$  for each spectator.

Integration over  $k$ in equation (\ref{eq:kxFF}) gives \begin{equation} \label{eq:xFF}
F(Q^2)  =  
\int_0^1\! dx \,e^{- \frac{ (1-x) Q^2 }{4 \kappa^2 x}},
\end{equation}
which is identical to the result obtained in the transverse impact representation after integration over 
$\mbf{b}_\perp$. The above expression for the form factor in the soft wall model is  valid for
relatively high $Q^2 $ since we have approximated the bulk-to-boundary propagator by its
asymptotic form. In the limit $Q^2 \gg \kappa^2$ we find
\begin{equation}
 F_\pi(Q^2) \to \frac{4 \kappa^2}{Q^2},
\end{equation}
the scaling result for a two-constituent meson at large $Q^2$ found
in Appendix D.

The pion structure function $q_\pi(x,Q^2)$ is computed by integrating the square of the
pion LFWF  (\ref{eq:LFWFkSW}) up to the scale $Q^2$
\begin{equation}
q_\pi(x,Q^2) = \int^{Q^2} \frac{d^2 \mbf{k}_\perp}{16 \pi^3} ~
\left\vert \psi_{\bar q q /\pi}(x, \mbf{k}_\perp)\right\vert^2.
\end{equation}
We find 
\begin{equation}
q_\pi(x,Q^2) = 1 - e^{-\frac{Q^2}{\kappa^2 x(1-x)}}.
\end{equation}
In the large $Q^2$ limit $q_\pi(x,Q^2 \to \infty) \equiv q_\pi(x) = 1$, which is the behavior of a strongly coupled theory found  in QCD(1+1)~\cite{Pauli:1985ps}. The result coincides with
(\ref{eq:sftau}) for $\tau = 2$. Identical results~\cite{Radyushkin:2006iz} are obtained for the pion in the hard wall model, 
$q(x) = 1$.

\end{document}